\documentclass[twocolumn]{aastex631}

\newcommand{\teff}{\tau_\mathrm{eff}}

\begin{document}
\title{\bf \large Constraints on the Evolution of the Ionizing Background and Ionizing Photon Mean Free Path at the End of Reionization}

\author[0000-0003-0821-3644]{Frederick B. Davies}
\affiliation{Max-Planck-Institut f\"{u}r Astronomie, K\"{o}nigstuhl 17, D-69117 Heidelberg, Germany}

\author[0000-0001-8582-7012]{Sarah E. I. Bosman}
\affiliation{Institute for Theoretical Physics, Heidelberg University, Philosophenweg 12, D–69120, Heidelberg, Germany}
\affiliation{Max-Planck-Institut f\"{u}r Astronomie, K\"{o}nigstuhl 17, D-69117 Heidelberg, Germany}

\author[0000-0002-2423-7905]{Prakash Gaikwad}
\affiliation{Max-Planck-Institut f\"{u}r Astronomie, K\"{o}nigstuhl 17, D-69117 Heidelberg, Germany}

\author[0000-0003-0294-8674]{Fahad Nasir}
\affiliation{Max-Planck-Institut f\"{u}r Astronomie, K\"{o}nigstuhl 17, D-69117 Heidelberg, Germany}

\author[0000-0002-7054-4332]{Joseph F. Hennawi}
\affiliation{Department of Physics, University of California, Santa Barbara, CA 93106, USA}
\affiliation{Leiden Observatory, Leiden University, Niels Bohrweg 2, 2333 CA Leiden, Netherlands}

\author[0000-0003-2344-263X]{George D. Becker}
\affiliation{Department of Physics and Astronomy, University of California, Riverside, CA, 92521, USA}

\author[0000-0001-8443-2393]{Martin G. Haehnelt}
\affiliation{Kavli Institute for Cosmology and Institute of Astronomy, Madingley Road, Cambridge, CB3 0HA, UK}

\author[0000-0003-3693-3091]{Valentina D'Odorico}
\affiliation{INAF-Osservatorio Astronomico di Trieste, Via Tiepolo 11, I-34143 Trieste, Italy}
\affiliation{Scuola Normale Superiore, Piazza dei Cavalieri 7, I-56126 Pisa, Italy}
\affiliation{IFPU-Institute for Fundamental Physics of the Universe, via Beirut 2, I-34151 Trieste, Italy}

\author[0000-0002-4314-021X]{Manuela Bischetti}
\affiliation{INAF–Osservatorio Astronomico di Trieste, Via G.B. Tiepolo, 11, I-34143 Trieste, Italy}
\affiliation{Dipartimento di Fisica, Sezione di Astronomia, Universitá di Trieste, via Tiepolo 11, 34143 Trieste, Italy}

\author[0000-0003-2895-6218]{Anna-Christina Eilers}\thanks{Pappalardo Fellow.}
\affiliation{MIT Kavli Institute for Astrophysics and Space Research, 77 Massachusetts Ave., Cambridge, MA 02139, USA}

\author[0000-0001-5211-1958]{Laura C. Keating}
\affiliation{Institute for Astronomy, University of Edinburgh, Blackford Hill, Edinburgh, EH9 3HJ, UK}

\author[0000-0001-5829-4716]{Girish Kulkarni}
\affiliation{Tata Institute of Fundamental Research, Homi Bhabha Road, Mumbai 400005, India}

\author[0000-0001-9372-4611]{Samuel Lai}
\affiliation{Research School of Astronomy and Astrophysics, Australian National University, Canberra, ACT 2611, Australia}

\author[0000-0002-5941-5214]{Chiara Mazzucchelli}
\affiliation{Instituto de Estudios Astrofísicos, Facultad de Ingeniería y Ciencias, Universidad Diego Portales, Avenida Ejercito Libertador 441, Santiago, Chile}

\author[0000-0002-4314-1810]{Yuxiang Qin}
\affiliation{School of Physics, University of Melbourne, Parkville, VIC 3010, Australia}
\affiliation{ARC Centre of Excellence for All Sky Astrophysics in 3 Dimensions (ASTRO 3D), Australia}

\author[0000-0001-5818-6838]{Sindhu Satyavolu}
\affiliation{Tata Institute of Fundamental Research, Homi Bhabha Road, Mumbai 400005, India}

\author[0000-0002-7633-431X]{Feige Wang}
\affiliation{Steward Observatory, University of Arizona, 933 N Cherry Avenue, Tucson, AZ 85721, USA}

\author[0000-0001-5287-4242]{Jinyi Yang}\thanks{Strittmatter Fellow.}
\affiliation{Steward Observatory, University of Arizona, 933 N Cherry Avenue, Tucson, AZ 85721, USA}

\author[0000-0003-3307-7525]{Yongda Zhu}
\affiliation{Department of Physics and Astronomy, University of California, Riverside, CA, 92521, USA}

\begin{abstract}
    The variations in Ly$\alpha$ forest opacity observed at $z>5.3$ between lines of sight to different background quasars are too strong to be caused by fluctuations in the density field alone. The leading hypothesis for the cause of this excess variance is a late, ongoing reionization process at redshifts below six. Another model proposes strong ionizing background fluctuations coupled to a short, spatially varying mean free path of ionizing photons, without explicitly invoking incomplete reionization. With recent observations suggesting a short mean free path at $z\sim6$, and a dramatic improvement in $z>5$ Ly$\alpha$ forest data quality, we revisit this latter possibility. Here we apply the likelihood-free inference technique of approximate Bayesian computation to jointly constrain the hydrogen photoionization rate $\Gamma_{\rm HI}$ and the mean free path of ionizing photons $\lambda_{\rm mfp}$ from the effective optical depth distributions at $z=5.0$--$6.1$ from XQR-30. We find that the observations are well-described by fluctuating mean free path models with average mean free paths that are consistent with the steep trend implied by independent measurements at $z\sim5$--$6$, with a concomitant rapid evolution of the photoionization rate. 
\end{abstract}


\section{Introduction}

The epoch of reionization reflects the cumulative photon output of the first generations of stars and galaxies in the Universe.
Determining the precise timing of reionization thus provides crucial clues for understanding structure formation in the first billion years of cosmic time. Observations of the cosmic microwave background suggest a characteristic epoch of $z\sim7$ \citep{Planck18}, and this rough midpoint is supported by constraints derived from the disappearance of Ly$\alpha$ emission from galaxies at $z>6$ (e.g. \citealt{Mason18,Mason19,Weinberger19,Hoag19,Jung20}, although see \citealt{Wold22}) and studies of the Ly$\alpha$ damping wing in the highest redshift quasars known at $z\gtrsim7$ (e.g. \citealt{Davies18b,Wang20,Yang20a,Greig22}).

The endpoint of reionization was originally estimated by the disappearance of widespread transmission in the Ly$\alpha$ forest \citep{Becker01,Fan06}, thought to be due to the onset of Gunn-Peterson absorption \citep{GP65} of neutral hydrogen in the intergalactic medium (IGM). That interpretation is complicated, however, by the concomitant decrease in the intensity of the ionizing background \citep{BH07a,WB11,Davies17,D'Aloisio18} which should lead to a mostly-opaque Ly$\alpha$ forest with strong sightline-to-sightline variations even without significantly neutral gas (e.g. \citealt{Lidz06b}). With the discovery of the giant Gunn-Peterson (GP) trough at $z\sim5.5$--$5.8$ towards the quasar ULAS J0148$+$0600 by \citet{Becker15}, however, existing models of the post-reionization IGM were no longer consistent with the distribution of Ly$\alpha$ forest opacity at $z\gtrsim5.6$. This inconsistency was confirmed by subsequent compilations of Ly$\alpha$ forest opacity measurements \citep{Bosman18,Eilers18} and has recently been carefully quantified to persist to even later times $z\sim5.3$ by \citet{Bosman22}.

Several models were put forward to explain the origin of these excess fluctuations, and in particular the existence of the giant GP trough from \citet{Becker15}. In ionization equilibrium, the Ly$\alpha$ forest opacity is directly connected to the residual neutral hydrogen fraction, which depends on the density of the gas, the strength of the ionizing background, and (through the recombination rate) the gas temperature. \citet{D'Aloisio15} suggested that relic temperature fluctuations from a late-ending, extended, and hot reionization process could lead to dramatic contrast in IGM temperatures at $z\lesssim6$, leading to strong variations in Ly$\alpha$ forest opacity due to the temperature dependence of the hydrogen recombination rate. Such models predict that opaque troughs correspond to large-scale galaxy overdensities \citep{Davies17b}, however, which is disfavored by galaxy surveys towards the largest GP troughs at $z\sim5.7$ \citep{Becker18,Kashino20,Christenson21}. \citet{Chardin15,Chardin17} showed that the strong fluctuations in the ionizing background would be a natural consequence of a bright and rare source population, i.e. luminous quasars (see also \citealt{Meiksin20}), but initial suggestions of an overabundant faint quasar population at $z>5$ \citep{Giallongo15} are now generally disfavored (e.g. \citealt{D'Aloisio17,Parsa18,Matsuoka18}). While interest in such quasar-dominated models has begun to resurface in light of the discovery of a substantial population of faint AGNs at $z>5$ by JWST surveys (e.g. \citealt{Kocevski23,Harikane23,Matthee23,Labbe23,Maiolino23}), due to their typically highly reddened nature it is still unknown whether these objects contribute substantially to the ionizing photon budget. \citet{DF16} explored the possibility that galaxy-sourced ionizing background fluctuations could be amplified by a coupling to the mean free path of ionizing photons (see also \citealt{D'Aloisio18}) following analytic arguments from \citet{McQuinn11}, but in order to match the broad distribution of Ly$\alpha$ forest opacity the \emph{average} mean free path would have to be a factor of a few shorter than the extrapolation from lower-redshift measurements \citep{Worseck14}. All of these models pre-supposed that reionization was more-or-less complete at $z\lesssim6$, and required adjusting their corresponding parameters to uncomfortable ends of the parameter space to be even qualitatively consistent with Ly$\alpha$ forest observations.

The most successful model thus far was proposed by \citet[][see also \citealt{ND20}]{Kulkarni19}, wherein reionization is \emph{not} finished by $z\sim6$, but instead ends at $z\lesssim5.5$. In this model, the large variations in the Ly$\alpha$ forest opacity at $z<6$ are explained by residual ``true" (i.e., mostly neutral) GP troughs and shadowing of the ionizing background by large-scale patches of neutral IGM. In particular, late reionization models most readily reproduce observations of rare, extremely large GP troughs \citep{Keating19} and so-called ``dark gaps'' \citep{Zhu21,Zhu22} even down to redshifts $z\lesssim5.5$. Such models, however, require careful tuning of the ionizing emissivity evolution to self-consistently reproduce both the late Ly$\alpha$ forest fluctuations and other reionization-epoch observables \citep{Kulkarni19,Keating19}, and the efficient moment-based radiative transfer method used in these works may suffer from numerical suppression of large-scale ionizing background fluctuations inside the ionized regions \citep{Wu21}. Nevertheless, this explanation has proven to be the most natural one so far, and has begun to deliver powerful constraints on the properties of the reionizing sources \citep{Choudhury21,Qin21}.

An important boundary condition to reionization is the strength of the ionizing background which permeates the Universe after the process is complete, and radiative transfer simulations generally predict a rapid rise in the amplitude of the ionizing background as the ionized bubbles overlap and the last neutral islands disappear (e.g. \citealt{Gnedin00}). 
Constraints on the hydrogen photoionization rate $\Gamma_{\rm HI}$ derived from Ly$\alpha$ forest observations suggest a rise by a factor of at least a few from $z\sim6$ to $z\sim5$ \citep{BH07a,WB11,Calverley11,Davies17}, followed by a much shallower evolution from $z\sim5$ to $z\sim2$
\citep{BB13}, which is potentially suggestive of this picture, although we note that the flattening could instead reflect changes in the nature of the gas responsible for the ionizing opacity \citep{Munoz16}. 
Precision measurements of $\Gamma_{\rm HI}$ at $z\sim5$--$6$ are complicated by the low mean transmission of the Ly$\alpha$ forest \citep{Eilers18,Bosman18} and a degeneracy with the relatively unknown thermal state of the IGM (e.g. \citealt{BB13}), but recent improvements in data quality \citep{Bosman22} and new constraints on the IGM thermal state at $z>5$ (e.g. \citealt{Walther19,Boera19,Gaikwad21}) suggest that we can now do much better.

Furthermore, the ionizing background at the very end of reionization provides a census of the ionizing photons being produced throughout the Universe which drove the process to completion, and introduces a strong boundary condition to models of the reionizing sources (e.g. \citealt{Bouwens15b}). Connecting the emissivity of ionizing photons $\epsilon_{\rm ion}$ to the photoionization rate $\Gamma_{\rm HI}$, however, requires an additional ingredient: the mean free path of ionizing photons, $\lambda_{\rm mfp}$. While the mean free path can be estimated from the distribution of neutral hydrogen systems at redshifts $z\lesssim4$ (e.g. \citealt{Rudie13,Prochaska14}), at higher redshifts the identification of individual absorption lines becomes more challenging. To overcome this difficulty, and to bypass potential uncertainties related to line blending and absorber clustering \citep{Prochaska14}, the mean free path can instead be measured \emph{directly} by stacking quasar spectra at their rest-frame Lyman limit \citep{Prochaska09}. Until recently, such measurements were limited to $z\lesssim5$ \citep{Prochaska09,OM12,Fumagalli13,Worseck14}, and so constraints on the ionizing emissivity were limited to a time well after the end of reionization \citep{BB13} or required an extrapolation of the mean free path evolution to earlier times \citep{D'Aloisio18}. 

Recently, \citet{Becker21} measured the mean free path of ionizing photons at $z\sim6$, improving the spectral stacking method by additionally modeling the impact of the quasar ionizing radiation on the line-of-sight Lyman-series and Lyman limit opacity. They measured a mean free path of $\lambda_{\rm mfp}=0.75^{+0.65}_{-0.45}$ proper Mpc, much shorter than the extrapolated value implied by the tight power-law evolution measured by \citet{Worseck14} at $z\lesssim5$. This mean free path is short enough to strongly affect the progression of reionization and place higher demands on the ionizing output from galaxies \citep{Cain21,Davies21}, and implicitly requires the structure of the gas to trace that of (cold) dark matter down to very small scales \citep{Emberson13,Park16,D'Aloisio20}. Notably, an interpolation between the $z=5.1$ and $z=6$ measurements of \citet{Becker21} now places the mean free path at the level required by the \citet{DF16} model to explain the Ly$\alpha$ forest fluctuations.

Here we revisit Ly$\alpha$ forest fluctuations in the context of the \citet{DF16} model to constrain parameters of the $z>5$ IGM, taking advantage of recent improvements in Ly$\alpha$ forest data quality and new constraints on the IGM thermal state \citep{Gaikwad21}. We apply a likelihood-free inference methodology based on \citet{Davies17} to the extended XQR-30 \citep{D'Odorico23} Ly$\alpha$ forest data set \citep{Bosman22} to constrain the average photoionization rate $\Gamma_{\rm HI}$ and the average mean free path of ionizing photons $\lambda_{\rm mfp}$ from $z=5.0$--$6.1$. We also assess the degree to which the data are consistent with a draw from the model. Finally, we discuss what our constraints imply for the evolution of the ionizing emissivity and the timing of the reionization epoch.

In this work we assume a $\Lambda$CDM cosmology with $h=0.685$, $\Omega_m=0.31$, $\Omega_b=0.049$, consistent with \emph{Planck} \citep{Planck18}. Distance units are comoving unless otherwise specified.

\section{Simulation Method} \label{sec:methods}

We simulate post-reionization Ly$\alpha$ forest fluctuations in a manner very similar to \citet{Davies17}. The general philosophy is to combine skewers through a cosmological hydrodynamical simulation, which can resolve scales small enough for a converged description of the Ly$\alpha$ forest, with skewers from a separate semi-numerical ionizing background simulation, which has a volume sufficiently large to sample the full distribution of large-scale background fluctuations. In the process, we unfortunately decouple the density of the gas responsible for Ly$\alpha$ absorption from the intensity of the ionizing radiation field. We will revisit this assumption in Section~\ref{sec:caveatemptor} and attempt to quantify the resulting biases.

\subsection{Nyx hydrodynamical simulation}

For the small-scale density, temperature, and velocity structure of the gas, we use a hydrodynamical simulation run with the \texttt{Nyx} code \citep{Almgren13}. The simulation we employ is 100$/h$ Mpc on a side and has 4096$^3$ dark matter particles and baryon grid cells, providing a box size and resolution adequate for converged Ly$\alpha$ forest statistics \citep{Lukic15,Onorbe18}. Our fiducial modeling used three redshift snapshots from $z=5$, $z=5.5$, and $z=6$; for redshifts between these values we use the overdensity field from the snapshot closest in redshift, and scale the physical densities by $(1+z)^3$. We extract 100,000 skewers of gas overdensity, temperature, and peculiar velocity from each snapshot starting from random locations in the simulation box and traversing in one of six random directions along grid axes ($\pm$x, y, z). Overdensities are converted into physical densities $\propto(1+z)^3$ according to the evolving redshift across the range of interest. Ly$\alpha$ forest spectra are computed by assuming ionization equilibrium to calculate the corresponding neutral hydrogen density and using the Voigt profile approximation of \citet{Tepper-Garcia06} to deposit Ly$\alpha$ opacity onto a 2\,km/s resolution grid.

One fundamental limitation of our Nyx simulation is that, by virtue of its optically-thin photoionization and photoheating rates from \citet{HM12}, reionization occurs very early ($z\sim15$), and with minimal heat injection ($\Delta T\sim10,000$\,K). This is in conflict with observational constraints on the timing of reionization (e.g. \citealt{Davies18b}) and theoretical models of the expected heating by ionization fronts sweeping through the IGM \citep{MR94,AH99,D'Aloisio18b}. As a result, at $z=5$--$6$ the IGM is colder, and the temperature-density relation steeper, than our theoretical understanding would otherwise suggest, with further evidence coming from recent constraints on the IGM thermal state at $z\sim5.5$ \citep{Gaikwad20}. The combination of these two effects will lead to overestimated recombination rates, and thus overestimated values of $\Gamma_{\rm HI}$ when matched to the observed Ly$\alpha$ forest opacity (e.g. \citealt{Bolton05,BH07a,BB13,D'Aloisio18}). To overcome this limitation of our hydrodynamical simulations, we adjust the IGM temperatures of our fiducial skewers to match a more realistic average post-reionization thermal state. We model the temperature evolution of the IGM following the analytic method from \citet{Davies17b} (see also \citealt{UptonSanderbeck16}) assuming an instantaneous reionization at $z_{\rm re}$ with a heat injection of $\Delta T=20,000$\,K (e.g. \citealt{D'Aloisio18b}). 

\begin{figure}
\begin{center}
\resizebox{8cm}{!}{\includegraphics[trim={1.0em 1em 1.0em 1em},clip]{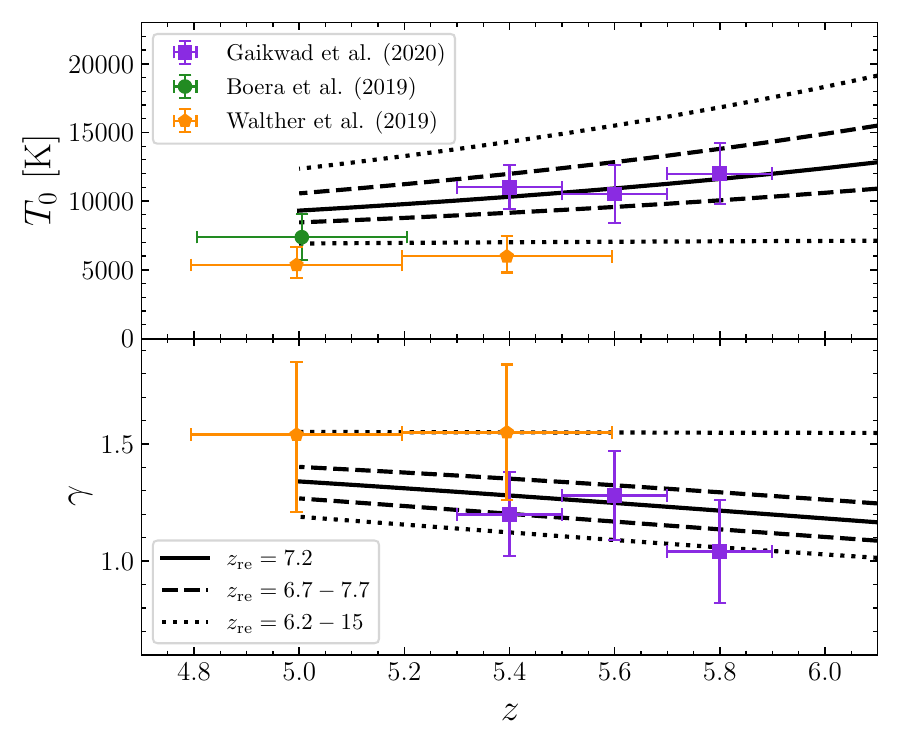}}
\end{center}
\caption{Evolution of the temperature at mean density in the IGM (top) and the slope of the temperature-density relation (bottom). The curves show the analytic models described in the text for our fiducial range of reionization heating models, while the data points show measurements from \citet{Walther19}, \citet{Boera19}, and \citet{Gaikwad20}.}
\label{fig:tempevol}
\end{figure}

In Figure~\ref{fig:tempevol} we show the resulting mean temperature $T_0$ and slope of the temperature-density relation $\gamma$ for a range of $z_{\rm re}$ compared to recent measurements of the IGM thermal state \citep{Walther19,Boera19,Gaikwad20}. From this comparison, we adopt $z_{\rm re}=7.2\pm0.5$ as our fiducial range of thermal models. However, we will also show the range of inferred $\Gamma_{\rm HI}$ values corresponding to more extreme thermal models: the cold, native Nyx temperatures corresponding to $z_{\rm re}\sim15$ and a very hot model with $z_{\rm re}=6.2$. By manually adjusting the simulated temperatures in post-processing, we neglect the effect of Jeans smoothing on the Ly$\alpha$ transmission, but this is likely a very small effect (see, e.g., \citealt{BB13,Kulkarni15}).

\subsection{Ionizing background fluctuations}

\begin{figure*}
\begin{center}
\resizebox{18cm}{!}{\includegraphics[trim={7.8em 0.2em 5.0em 2.8em},clip]{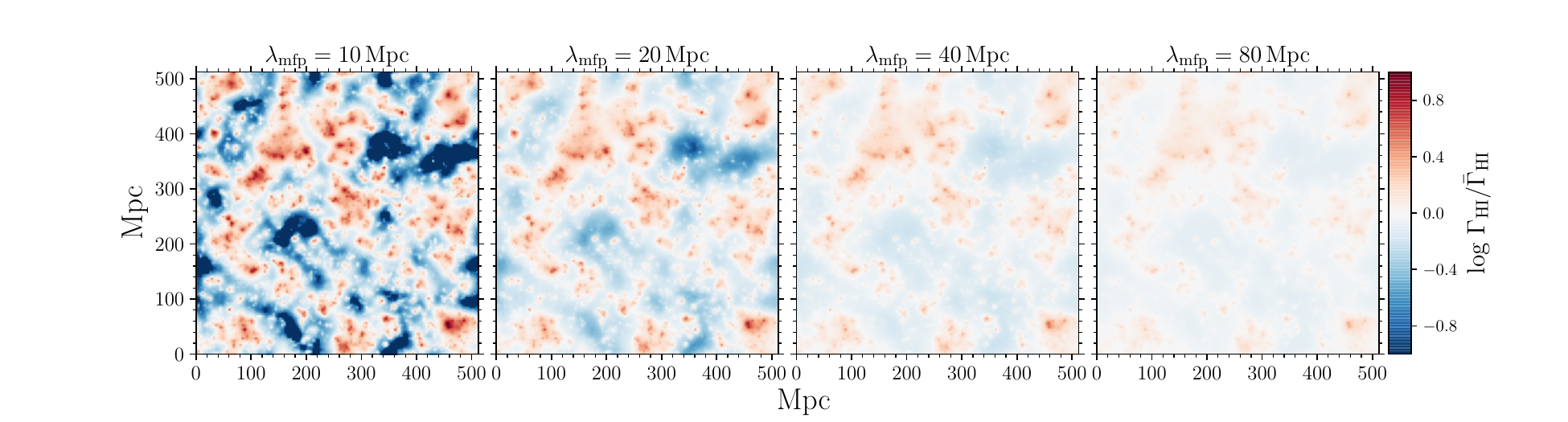}}
\end{center}
\caption{Slices through a subset of our fluctuating $\Gamma_{\rm HI}$ fields, 4\,Mpc-thick and 512 Mpc on a side, showing a representative range of $\lambda_{\rm mfp}$ values of 10, 20, 40, and 80\,Mpc from left to right.}
\label{fig:uvbmap}
\end{figure*}

The volume of the Nyx simulation described above is not entirely sufficient to capture the large-scale ionizing background fluctuations characteristic of the post-reionization IGM, where box sizes of at least a few hundred Mpc are required (e.g. \citealt{DF16,D'Aloisio18}). For our fiducial model, we decided to create a much larger, but independent, cosmological volume within which to simulate the ionizing background -- the consequences and biases of this approach will be discussed later.

We computed large-scale simulations of ionizing background fluctuations using the method of \citet{DF16}, built off of the semi-numerical framework of the \texttt{21cmFAST} code from \citet{Mesinger11}. Cosmological initial conditions were generated on a $4096^3$ grid in a volume 512 comoving Mpc on a side, then evolved to $z=5.5$ onto a coarser $512^3$ grid via the Zel'dovich approximation \citep{Zel'dovich70}. Halos were instantiated from the 4096$^3$ grid via the excursion set halo-filtering approach of \citet{MF07} down to a minimum halo mass $M_{h,{\rm min}}=2\times10^{9}$\,$M_\odot$, and shifted to their evolved positions at $z=5.5$ via the Zel'dovich displacement field evaluated on the $512^3$ grid at their initial positions from linear theory, a procedure which produces realistic halo clustering down to $\sim$Mpc scales \citep{MF07}. Ultraviolet (UV) luminosities of galaxies corresponding to each halo were assigned by abundance matching to the \citet{Bouwens15} UV luminosity functions, resulting in UV magnitudes ranging from $-12.6 \gtrsim M_{\rm UV} \gtrsim -23.3$. For simplicity, we assume that the ionizing luminosity of each galaxy is proportional to its UV luminosity, and leave the constant of proportionality as a free parameter.

We then compute the ionizing background by a brute force radiative transfer algorithm applied to a coarser, $128^3$ grid, corresponding to cells 4 Mpc on a side. The ionizing luminosities of the galaxies are deposited onto the coarse grid, and then the ionizing radiation intensity in cell $i$ is computed via
\begin{equation}
    J_{\nu,i} = J_{\nu,i}^{\rm local} + \sum_{j \neq i} \frac{L_{\nu,j}}{(4\pi)^2 d_{ij}^2} e^{-\tau_{ij}(\nu)},
\end{equation}
where $J_{\nu,i}^{\rm local}$ is the intensity due to local sources, $d_{ij}$ is the distance between cells $i$ and $j$, and $\tau_{ij}$ is the ionizing photon optical depth between $i$ and $j$. The sum over all other cells $j$ is performed assuming locally periodic boundary conditions, i.e. the cell $i$ ``sees'' the other cells $j$ within a volume of the same size as the full box but centered on cell $i$. The local source intensity is computed assuming a uniform ionizing emissivity $\epsilon_{\nu,i}$ inside the cell,
\begin{equation}
    J_{\nu,i}^{\rm local} = \frac{\epsilon_{\nu,i}\lambda_{\nu,i}}{4\pi}\left(1-e^{-0.72l/\lambda_{\nu,i}}\right) ,
\end{equation}
where $\lambda_{\nu,i}$ is the mean free path of cell $i$ at frequency $\nu$, $l$ is the side length of the cell, and the constant factor of 0.72 was found to have good resolution convergence properties in \citet{DF16}. The optical depth between cells is computed by integrating the opacity $\kappa = d\tau/dr$ on the 128$^3$ grid,
\begin{equation}
    \tau_{ij}(\nu) = \int_i^j \kappa(\nu,\Gamma,\Delta) dr,
\end{equation}
where $\kappa$ is assumed to vary with the local photoionization rate $\Gamma$ and overdensity relative to the cosmic mean $\Delta=\rho/\bar{\rho}$ according to
\begin{equation}\label{eqn:mfplaw}
    \kappa = \kappa_0 \times \Delta \left(\frac{\nu}{\nu_{\rm HI}}\right)^{-0.9} \left(\frac{\Gamma}{\Gamma_0}\right)^{-2/3},
\end{equation}
where $\kappa_0$ is the opacity of the IGM at the ionizing edge of hydrogen $\nu_{\rm HI}$ and at a reference photoionization rate $\Gamma_0$, and we adopt the same power-law dependencies on $\Delta$ and $\nu$ as \citet{DF16}. Finally, the photoionization rate is calculated by integrating over $J_\nu$,
\begin{eqnarray}
    \Gamma_{{\rm HI},i} &=& 4\pi\int_{\nu_{\rm HI}}^{\infty} \frac{J_{\nu,i}}{h\nu} \sigma_{\rm HI}(\nu) d\nu \nonumber \\
    &\propto& \frac{J_{\bar{\nu},i}}{h\bar{\nu}}\sigma_{\rm HI}(\bar{\nu})\Delta\nu, 
\end{eqnarray}
where $\sigma_{\rm HI}$ is the hydrogen photoionization cross-section (from \citealt{Verner96}). In practice, for computational efficiency we perform the calculation with a single ionizing photon frequency $h \bar{\nu} \approx 17.9$\,eV, which was found to produce very similar $\Gamma_{\rm HI}$ fluctuations to a more comprehensive multi-frequency calculation \citep{DF16}. 

We computed fluctuating $\Gamma_{\rm HI}$ fields for 14 different average mean free path values $\lambda_{\rm mfp} = \kappa_0^{-1}= 5$, 6, 8, 10, 15, 20, 25, 30, 40, 50, 60, 80, 100, 150\,(comoving) Mpc at a fixed redshift $z=5.5$. This relatively coarse sampling was necessitated by the computational inefficiency of the \citet{DF16} method; see \citet{Gaikwad23} for a recent implementation of a more efficient algorithm.
We first computed the $\Gamma_{\rm HI}$ field assuming a uniform $\kappa=\kappa_0(\bar{\nu}/\nu_{\rm HI})^{-0.9}$, and then determine a normalization factor (corresponding to a combination of an ionizing emissivity normalization and a correction for the monochromatic approximation) to initialize the mean $\Gamma_{\rm HI}$ to the value corresponding to the prescribed ionizing opacity, i.e. $\Gamma_0$ in equation~(\ref{eqn:mfplaw}).  
In the following, we will use only the \emph{relative} fluctuations in $\Gamma$, i.e. the relative fluctuating field $\tilde{\Gamma}_{\rm HI}\equiv\Gamma_{\rm HI}/\bar{\Gamma}_{\rm HI}$, so the exact value of $\Gamma_0$ is unimportant. Note that due to the dependence of $\kappa$ on $\Gamma_{\rm HI}$, the calculation must be iterated many times to achieve self-consistency. We iterate the calculation until the average change in $\Gamma_{\rm HI}$ falls below $0.1\%$, requiring $\gtrsim20$ iterations for $\lambda_{\rm mfp}\lesssim10$\,Mpc and $\sim5$ iterations for $\lambda_{\rm mfp}\gtrsim60$\,Mpc. 

In Figure~\ref{fig:uvbmap}, we show a slice through four of the fluctuating ionizing background fields with $\lambda_{\rm mfp}=10$--$80$\,Mpc. The spatial structure of the fluctuations is very similar, driven by the large-scale clustering of the ionizing sources, but the fluctuations are stronger at shorter average mean free path. We take advantage of this coherence and linearly interpolate in $\log{\tilde{\Gamma}}$ between adjacent models to produce skewers at arbitrary average mean free path from 5 to 150\,Mpc. 

As discussed in the following Section, our inference procedure requires many millions of simulated Ly$\alpha$ forest spectra. To reduce the computational requirements for producing the mock spectra, we adopt an approach analogous to the fluctuating Gunn-Peterson approximation \citep{Weinberg97} to both adjust the mean $\Gamma_{\rm HI}$ and introduce the fluctuating $\Gamma_{\rm HI}$ field. We first compute an initial set of 100,000 Ly$\alpha$ forest spectra from the Nyx simulation skewers with a uniform $\Gamma_{\rm HI}=\Gamma_{\rm init}$, where we set $\Gamma_{\rm init}$ to the midpoint of the redshift-dependent uniform prior on $\Gamma_{\rm HI}$ used during the inference procedure (see \S~\ref{sec:constraints}). We then introduce $\Gamma_{\rm HI}$ fluctuations by re-scaling the pixel optical depths along the skewer,
\begin{equation}
    \tau_\alpha(z_{{\rm Ly\alpha},i}) \propto \left(\frac{\Gamma_{\rm new}\times\tilde{\Gamma}_{\rm HI}(z_i)}{\Gamma_{\rm init}}\right)^{-1},
\end{equation}
where $\Gamma_{\rm new}$ is the new average value of $\Gamma_{\rm HI}$ and $\tilde{\Gamma}_{\rm HI}$ is evaluated at $z_i$ without taking into account redshift-space distortions. We find that this procedure is $\sim50$ times faster than a more accurate re-evaluation of the spectra from the neutral hydrogen distribution, without substantially affecting our main results.

While the ionizing background is inherently fluctuating in our simulations, references to ``$\Gamma_{\rm HI}$'' in the rest of the text, particularly in the context of our constraints from observations, will represent the volume-averaged value of $\Gamma_{\rm HI}$.

\subsubsection{Self-consistent background fluctuations} \label{sec:selfcon}

For comparison and bias assessment purposes, we also produced a similar suite of simulations of ionizing background fluctuations using the halo catalog of the $100$\,Mpc$/h$ Nyx simulation with the same set of average mean free path values. We abundance-matched dark matter halos down to the same minimum mass, $M_{\rm min}=2\times10^9$\,$M_\odot$, but we note that due to the coarse force resolution of the uniform grid particle-mesh scheme used to evolve the dark matter particles, the number of halos below $M_h\sim10^{10}$\,$M_\odot$ is significantly underestimated (e.g. \citealt{Lukic07,Almgren13}). As a result, the UV magnitude range is somewhat restricted compared to our larger volume, $-13.9\gtrsim M_{UV} \gtrsim -22.7$. 

For these complementary simulations, we computed the background fluctuations on a $64^3$ grid, corresponding to a spatial resolution of $\sim2$ Mpc. These simulations tend to have weaker large-scale fluctuations than the fiducial ones for relatively short mean free paths, but for long mean free paths they exhibit stronger large-scale fluctuations. The former is due to the suppression of large-scale modes in the density field due to the smaller volume, the latter is due to a limitation of the algorithm used to compute the radiation field. Specifically, the size of the ``local volume'' seen by any cell sets an upper limit to the effective mean free path of roughly half the box size, or in this case $\sim73$\,Mpc. 

We will refer to this suite of simulations as the ``self-consistent'' model, as it provides ionizing background fluctuations sourced by the same underlying density field as the Ly$\alpha$ forest. As mentioned above, the fiducial model relies on an independent volume for ionizing background fluctuations, which inherently decouples them from the density field. We will compare the constraints obtained in the fiducial model to the self-consistent model, as well as a de-correlated version of the self-consistent model, where the ionizing background fluctuations are (as in the fiducial model) drawn from a random region in the $100$\,Mpc$/h$ volume.

\section{Likelihood-free inference with Approximate Bayesian Computation}

\begin{figure*}
\begin{center}
\resizebox{18cm}{!}{\includegraphics[trim={1em 1em 1em 1em},clip]{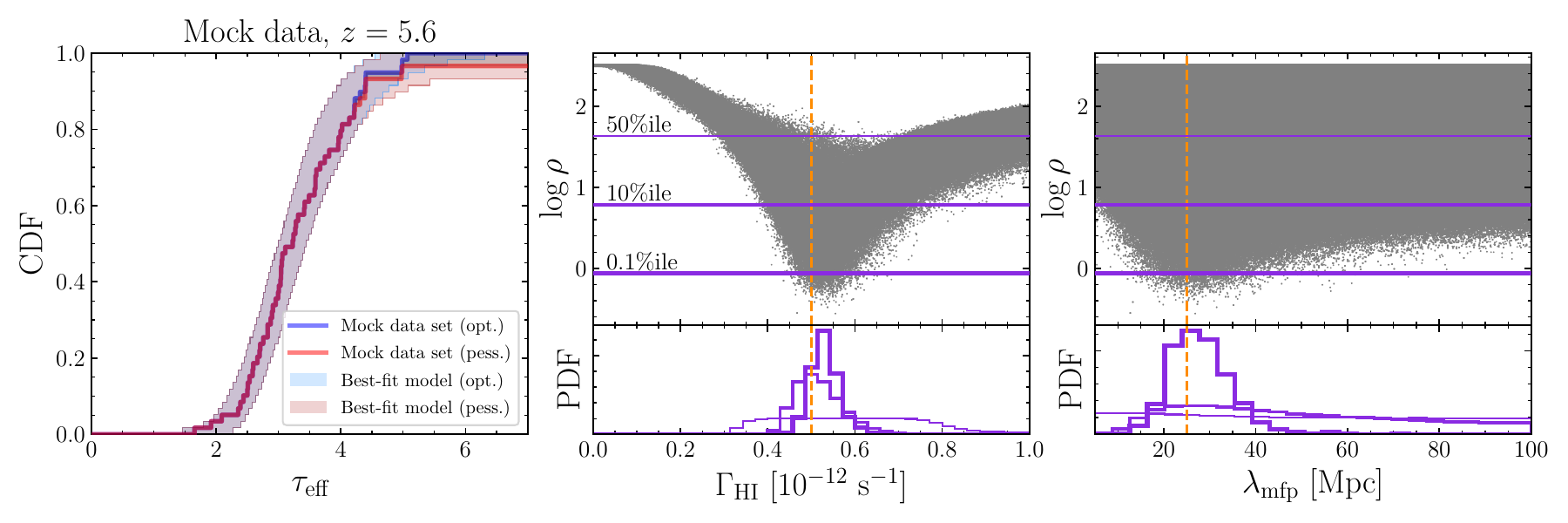}}
\end{center}
\caption{Demonstration of ABC on a mock data set. The blue and red curves in the left panel correspond to the cumulative distribution functions of Ly$\alpha$ forest effective optical depth from a mock XQR-30 data set, with non-detections treated as optimistic ($F=2\times\sigma_F$) or pessimistic ($F=0$), respectively, following \citet{Bosman18,Bosman22}. The shaded regions correspond to the central 95\% of the scatter of additional mock data sets with $\Gamma_{\rm HI}$ and $\lambda_{\rm mfp}$ set to their mean posterior estimates. The grey points in the upper halves of the middle and right panels show the distance metric $\rho$ (equation~\ref{eqn:abcdist}) computed from 1,000,000 mock data sets. The horizontal lines show three different $\rho$ thresholds below which lie 50\%, 10\%, and 0.1\% of the mock data samples from top to bottom. The corresponding posterior PDFs on $\Gamma_{\rm HI}$ and $\lambda_{\rm mfp}$ are shown in the lower panels, with the true values indicated by the vertical dashed lines.}
\label{fig:abcexample}
\end{figure*}

Armed with the simulation framework described above, we aim to constrain two parameters: $\Gamma_{\rm HI}$ and $\lambda_{\rm mfp}$. The distribution of Ly$\alpha$ forest opacity at $z>5$ is not well-described by Gaussian statistics (e.g. \citealt{Bosman18}) that would typically be adopted in Bayesian parameter inference, and at $z\gtrsim5.6$ one must also sensibly account for non-detections that represent the most constraining GP troughs. We adopt a likelihood-free inference technique known as Approximate Bayesian Computation or ABC \citep{Rubin84,Tavare97,Pritchard99} to overcome these challenges with a principled approach\footnote{We note that more sophisticated and efficient likelihood-free inference methods aided by machine learning have recently been developed (e.g. \citealt{Alsing18,Cole22}), but we adopt ABC for conceptual simplicity.}. Our method is very similar to that described in \citet{Davies17}, but with some important differences that we describe below.

The approximation fundamental to the ``approximate'' nature of ABC is in the definition of the likelihood of the data vector ${\bf d}$ given the model parameters $\theta$, $p({\bf d}|\theta)$. In ABC, the likelihood is approximated by (e.g. \citealt{Marin12})
\begin{equation}
    p({\bf d}|\theta) \approx p(\rho({\bf d},{\bf x}(\theta)) < \epsilon|\theta) \approx p(\rho({\bf s_d},{\bf s_x}(\theta))|\theta)
\end{equation}
where $\rho$ is a (nearly arbitrary) distance metric between the data and a mock data draw ${\bf x}$ and $\epsilon$ is a distance threshold below which the data and mock data are deemed ``similar enough''. As $\epsilon$ approaches zero, the approximate posterior PDF will converge towards the true posterior PDF (e.g. \citealt{Blum10}). Typically the raw data are first transformed into a lower dimensional summary statistic ${\bf s}({\bf d})\equiv{\bf s_d}$, and the same procedure is applied to make mock observations of the summary statistic ${\bf s}({\bf x})\equiv{\bf s_x}$. The ABC procedure involves computing many such mock data sets, with parameter values $\theta$ drawn from the prior $p(\theta)$, and selecting some number of samples with distances below some threshold $\epsilon$ -- those samples then (approximately) represent samples from the posterior PDF $p(\theta|{\bf d})$. As discussed below, the threshold $\epsilon$ is typically chosen such that one retains a given (small) fraction of the parameter samples. 

There is substantial freedom in choosing both the distance metric $\rho$ and the summary statistic ${\bf s}$. We choose to summarize the Ly$\alpha$ forest data by their $\Delta z=0.1$-binned effective optical depths ${\bf s} = -\ln{\langle\exp{-\tau_\alpha}\rangle} = \tau_{\rm eff}$, and compute the Euclidean ($L_2$ norm) distance between the \emph{rank-order} set of observed and mock $\tau_{\rm eff}$ (see also \citealt{Davies17}). Specifically, we have
\begin{equation}\label{eqn:abcdist}
    \rho({\bf s_d},{\bf s_x}) = \sqrt{\sum_i^{N_{\rm q}} (\tau_{{\rm eff},i}^{\rm obs}-\tau_{{\rm eff},i}^{\rm mock})^2},
\end{equation}
where $\tau_{{\rm eff},i}^{\rm obs}i$ and $\tau_{{\rm eff},i}^{\rm mock}$ are the $i$th highest $\tau_{\rm eff}$ values in the set of $N_{\rm q}$ observed and mock Ly$\alpha$ forest spectra, respectively. While the transformation to effective optical depth provides sensitivity to strongly absorbed regions of the Ly$\alpha$ forest, observational noise can lead to fully opaque GP troughs with mean transmission below zero, leading to undefined $\tau_{\rm eff}$. In both the observed data and mock data, we set $\tau_{\rm eff}=-\ln{(2\times\sigma_F)}$ if the mean transmission is below twice the statistical error $\sigma_F$; a common transformation in the $z>5$ Ly$\alpha$ forest literature \citep{Becker15,Eilers18,Bosman18,Bosman22}. 

In Figure~\ref{fig:abcexample}, we demonstrate an example of the procedure described above applied to a mock Ly$\alpha$ forest data set at $z=5.6$. We assume true values of $\Gamma_{\rm HI}=5\times10^{-13}$\,s$^{-1}$ and $\lambda_{\rm mfp}=25$\,Mpc, and a data set consisting of $50$ quasar sightlines with a statistical error $\sigma_F=0.01$ on the $\Delta z=0.1$-binned transmitted flux and a continuum error $\sigma_C=0.1$. The cumulative distribution function (CDF) of this ``observed'' data set is shown by the thick curves in the left panel of Figure~\ref{fig:abcexample}, where the left (blue) and right (red) curves display undetected transmission with $F<2\sigma_F$ as $\tau_{\rm eff}=-\ln{(2\sigma_F)}$ or $\tau_{\rm eff}=\infty$, respectively, similar to \citet{Bosman18,Bosman22}. We draw 1,000,000 values of $\theta=\{\Gamma_{\rm HI},\lambda_{\rm mfp}\}$ assuming a uniform prior from $\Gamma_{\rm HI}=0$--$10^{-12}$\,s$^{-1}$ and $\lambda_{\rm mfp}=5$--$150$\,Mpc, and for each $\theta$ we compute a mock data set including the same noise properties as the observed one. The resulting distances $\rho({\bf s_d},{\bf s_x})$ vs. $\Gamma_{\rm HI}$ and $\lambda_{\rm mfp}$ are shown in the (upper) middle and right panels of Figure~\ref{fig:abcexample}, respectively. We then choose a series of distance thresholds $\epsilon$ such that 50\%/10\%/0.1\% of the mock data sets have $\rho({\bf s_d},{\bf s_x}) < \epsilon$, shown by the horizontal lines, and retain the corresponding $\theta$ values as samples from the posterior PDFs in the lower panels. The blue and red shaded regions curves in the left panel of Figure~\ref{fig:abcexample} show the central 95\% of CDFs computed from mock data sets adopting the lowest $\epsilon$ posterior mean $\Gamma_{\rm HI}$ and $\lambda_{\rm mfp}$, demonstrating good consistency with the ``observed'' one.

\begin{figure*}
\begin{center}
\resizebox{18cm}{!}{\includegraphics[trim={1.0em 1em 1.0em 1em},clip]{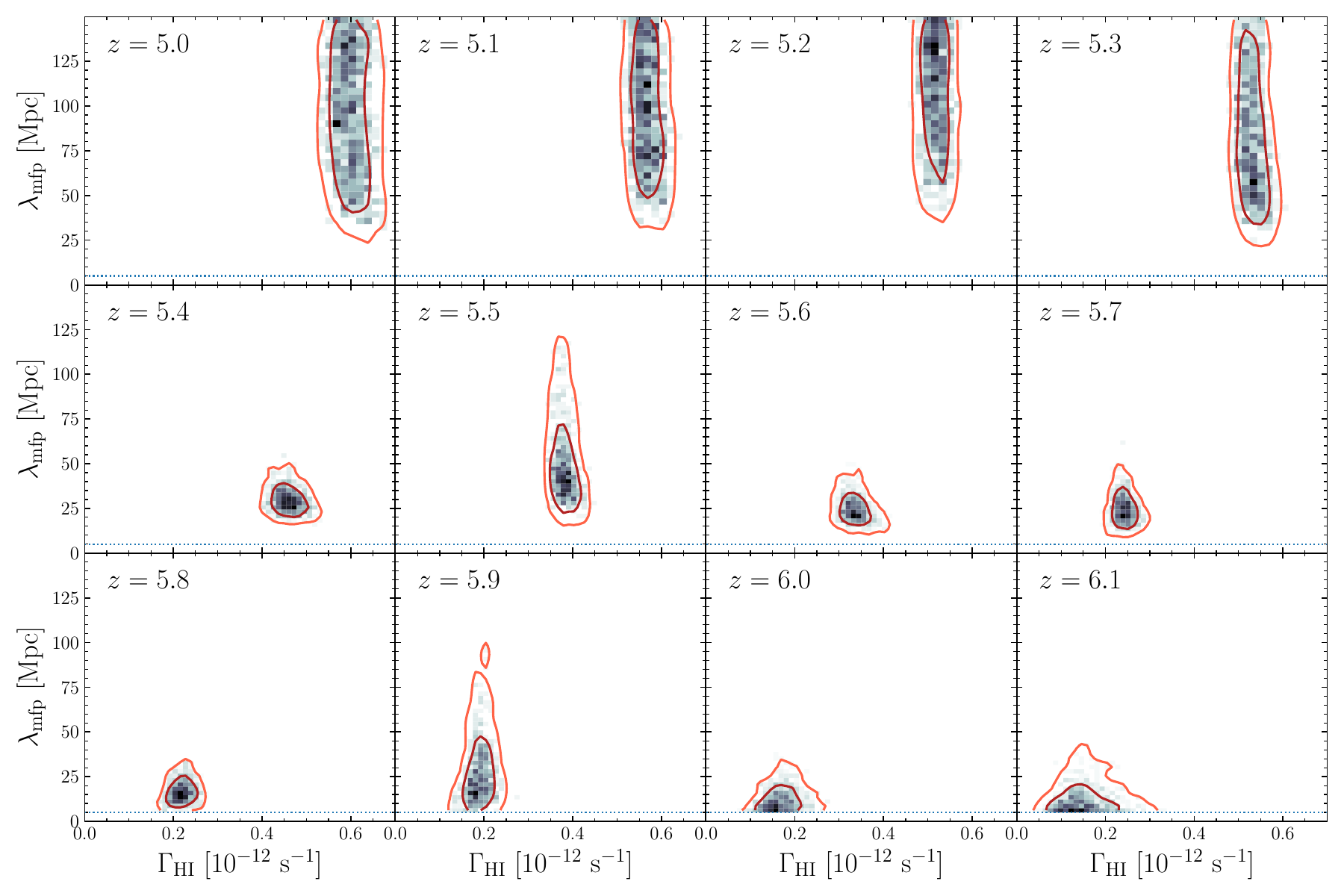}}
\end{center}
\caption{Joint posterior PDF of $\Gamma_{\rm HI}$ and $\lambda_{\rm mfp}$ at $z=5.0$--$6.1$. The inner and outer contours contain 68\% and 95\% of the distribution, respectively. The blue dashed line shows the lower edge of the $\lambda_{\rm mfp}$ prior.}
\label{fig:abcpdfs}
\end{figure*}

\section{Constraints on $\Gamma_{\rm HI}$ and $\lambda_{\rm mfp}$ from the extended XQR-30 Data Set}\label{sec:constraints}

\begin{figure}
\begin{center}
\resizebox{8cm}{!}{\includegraphics[trim={1.0em 1em 1.0em 1em},clip]{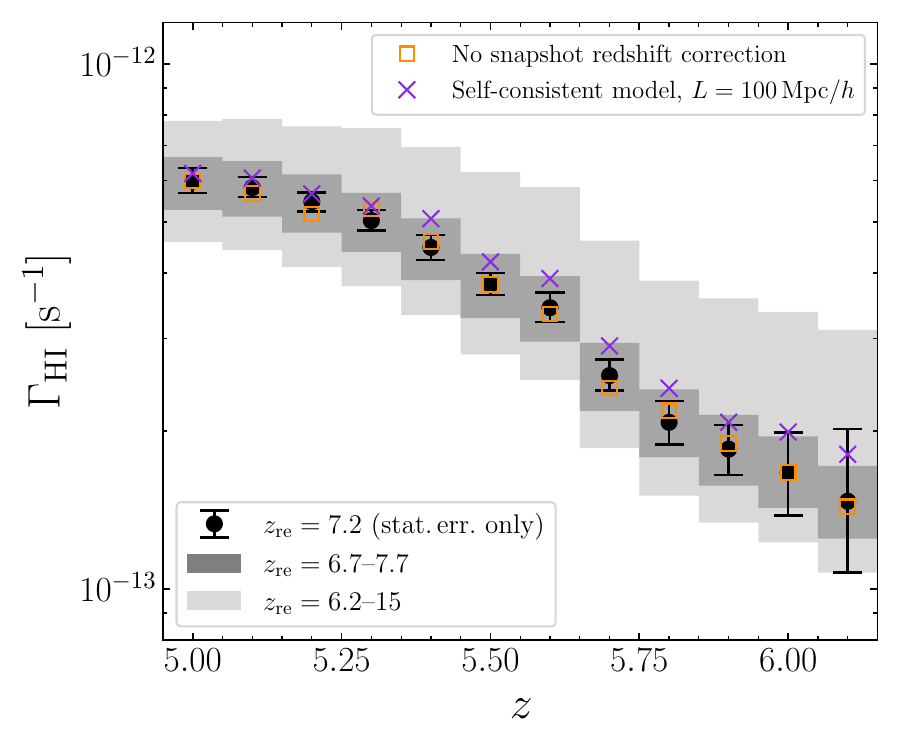}}
\end{center}
\caption{Posterior medians (black circles) and central 68\% credible intervals (black thin error bars) on $\Gamma_{\rm HI}$ from the XQR-30 data set assuming $z_{\rm re}=7.2$. The dark grey shaded region shows the deviation of the posterior medians for $z_{\rm re}=6.7$ and $z_{\rm re}=7.7$, while the light grey shaded region shows the range from more extreme thermal models with $z_{\rm re}=6.2$ and $z_{\rm re}\sim15$. The open orange points show the constraints without the correction for the coarse redshift snapshot sampling (see Appendix~\ref{sec:appendix_snaps}). The purple crosses show the posterior medians from the self-consistent model in the $L=100$\,Mpc$/h$ hydrodynamical simulation volume, see \S~\ref{sec:selfcon}.}
\label{fig:abcghi}
\end{figure}

\begin{figure*}
\begin{center}
\resizebox{8.5cm}{!}{\includegraphics[trim={1.0em 1em 1.0em 1em},clip]{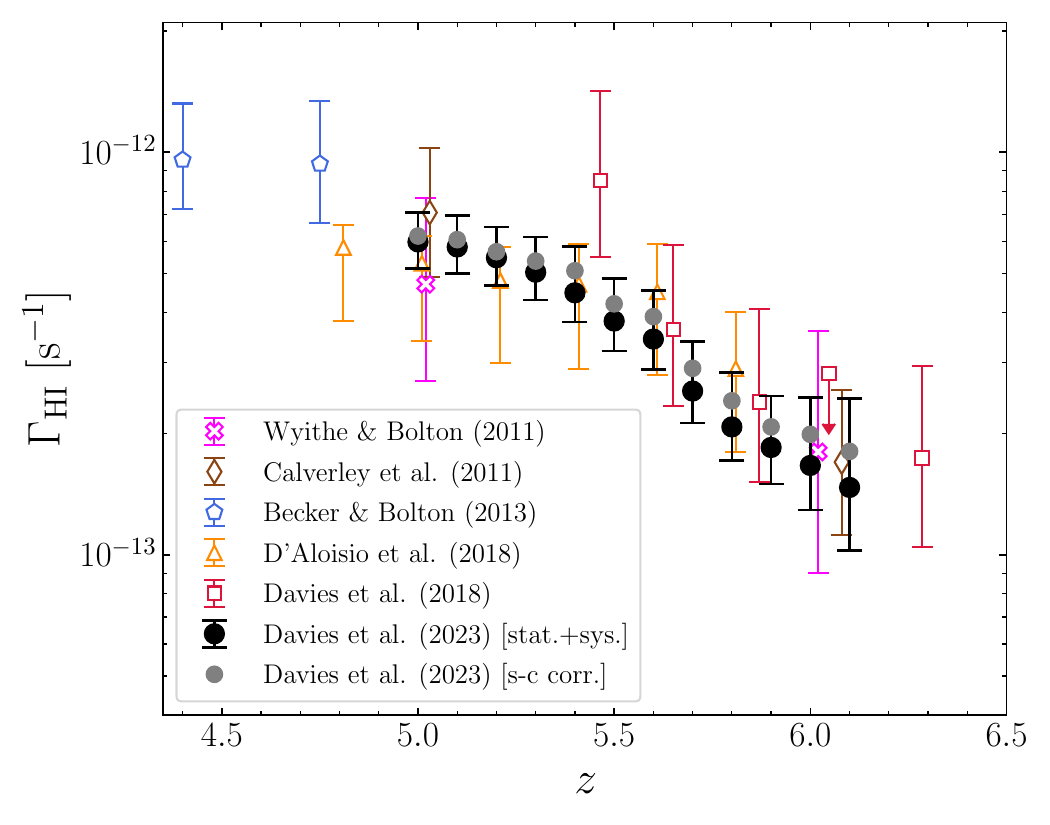}}
\resizebox{8.5cm}{!}{\includegraphics[trim={1.0em 1em 1.0em 1em},clip]{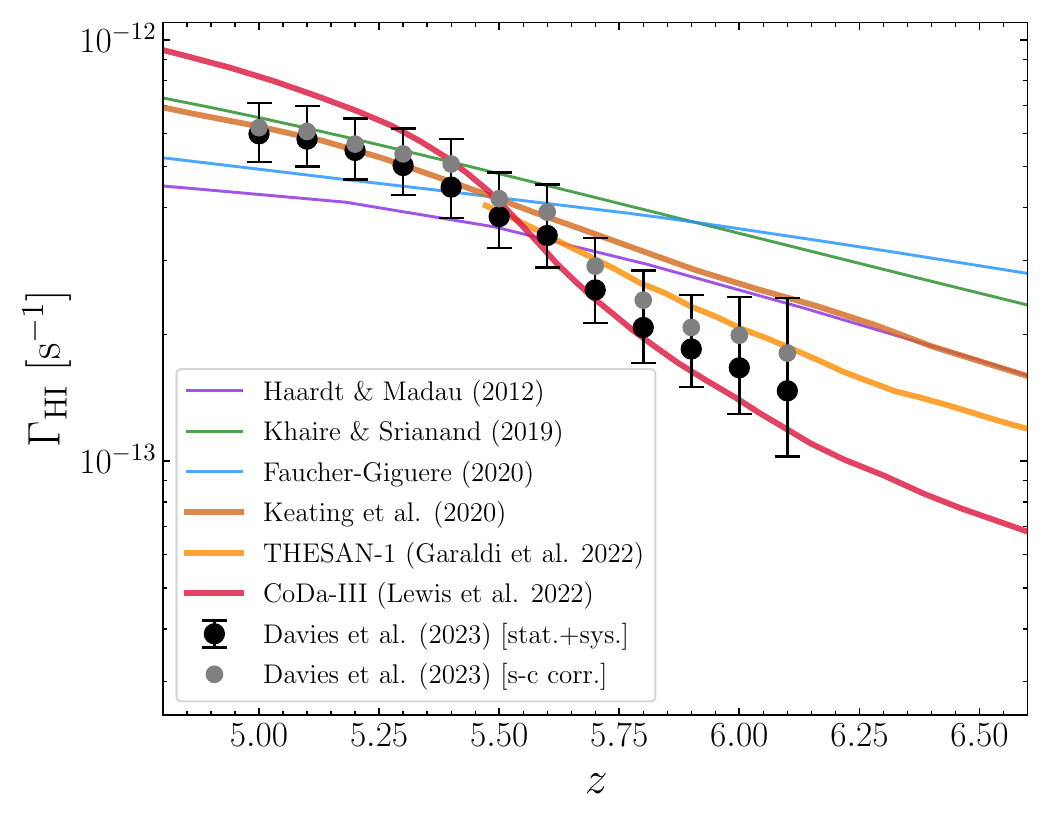}}
\end{center}
\caption{Left: Comparison of our $\Gamma_{\rm HI}$ constraints (black points with error bars, including statistical and systematic uncertainty) to previous measurements from the literature: \citet{Calverley11} (brown diamonds; quasar proximity zone profiles); \citet{WB11}, \citet{BB13}, and \citet{D'Aloisio18} (pink crosses, blue pentagons, and orange triangles; mean Ly$\alpha$ forest transmission); \citet{Davies17} (red squares; Ly$\alpha$ and Ly$\beta$ transmission spikes). No corrections have been made for differences in cosmology or assumptions of the IGM thermal state between these works. The grey points show our constraints with an approximate correction for a bias due to the lack of self-consistency. Right: Comparison to theoretical models of the ionizing background (curves), computed from 1D cosmological radiative transfer calculations by \citet{HM12} (purple), \citet{KS19} (green), \citet{FG20} (blue) and 3D radiation-hydrodynamic cosmological simulations by \citet{Keating19} (brown), \citet{Garaldi22} (orange; THESAN), and \citet{Lewis22} (red; CoDa-III).}
\label{fig:abcghi2}
\end{figure*}

\begin{figure}
\begin{center}
\resizebox{8cm}{!}{\includegraphics[trim={1.0em 1em 0.0em 1.0em},clip]{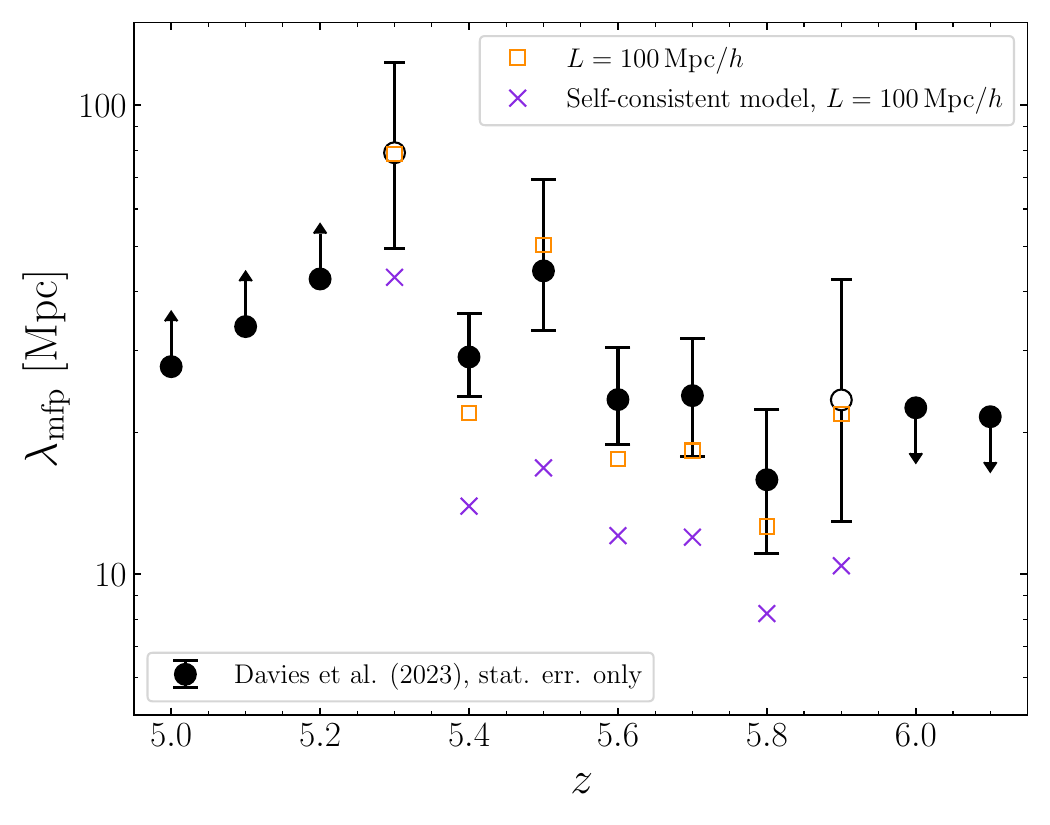}}
\end{center}
\caption{Posterior medians (black circles) and central 68\% credible intervals (black thin error bars) on $\lambda_{\rm mfp}$ from the XQR-30 data set, with open circles corresponding to marginal constraints and arrows corresponding to $2\sigma$ limits (see text for details). Open orange points show the posterior medians from the 100\,Mpc$/h$ simulation with uncorrelated density and ionizing background, while the purple crosses show the posterior medians from the fully self-consistent simulation.}
\label{fig:abcmfp}
\end{figure}

\begin{figure*}
\begin{center}
\resizebox{8.5cm}{!}{\includegraphics[trim={1.0em 1em 1.0em 1em},clip]{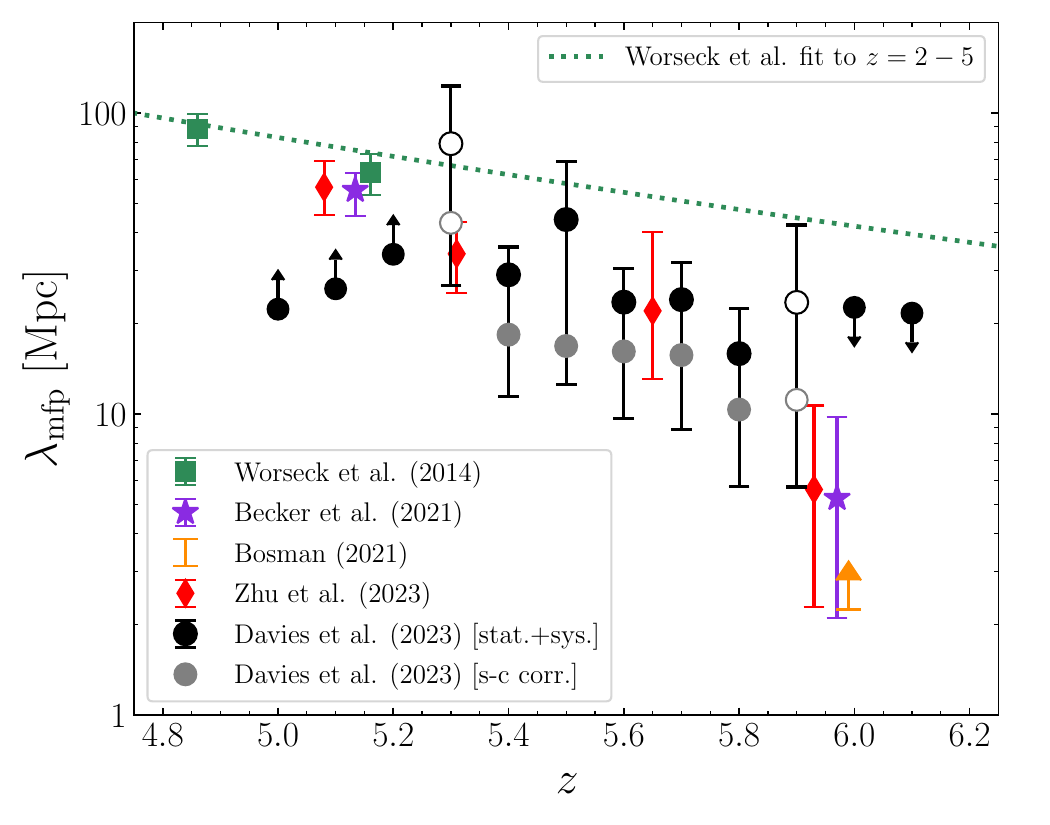}}
\resizebox{8.5cm}{!}{\includegraphics[trim={1.0em 1em 1.0em 1em},clip]{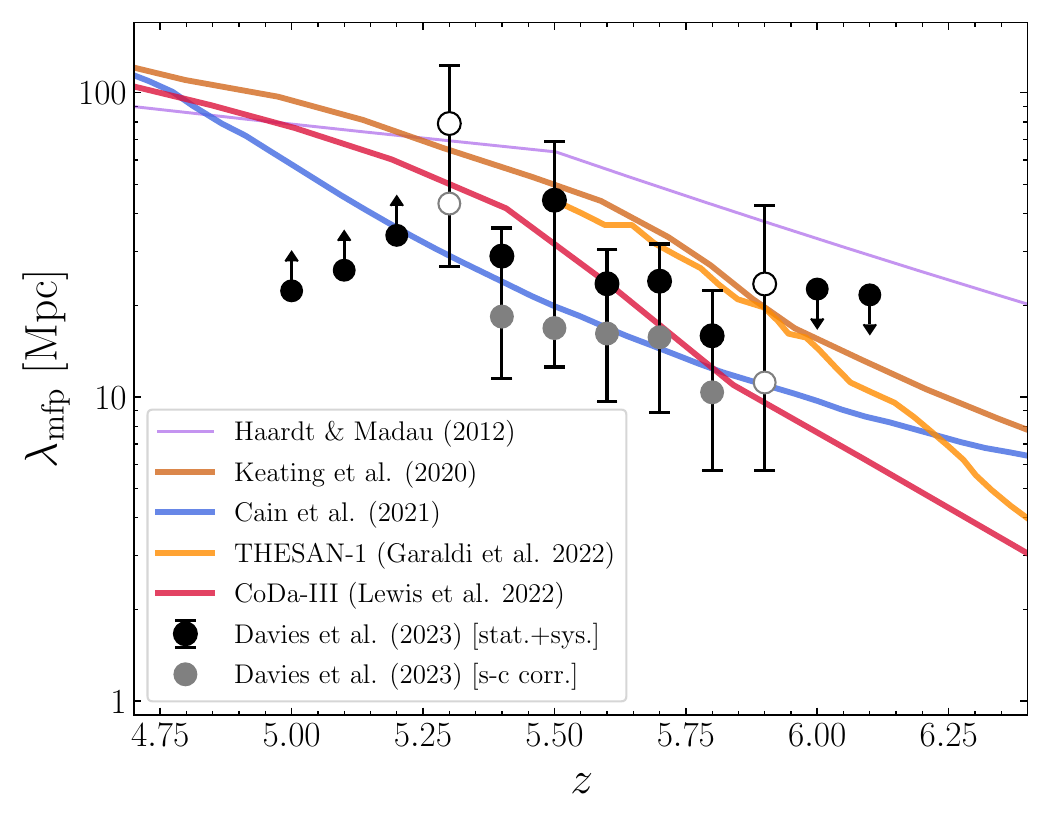}}
\end{center}
\caption{Left: Our constraints on $\lambda_{\rm mfp}$ compared to observations are shown by the black points, while the grey points show the constraints shifted to approximately correct for a bias due to the lack of self-consistency between the density field and ionizing background. Measurements from \citet{Worseck14} (green squares), \citet{Becker21} (purple pentagons), \citet{Zhu23} (red diamonds), and the lower limit from \citet{Bosman21MFP} (orange arrow) are shown for comparison. The dotted line shows the extrapolated fit to the full redshift range of measurements from \citet{Worseck14}. Right: Comparison to the mean free path in theoretical models, either prescribed by extrapolation of empirical constraints on the column density distribution of hydrogen absorbers \citep[][purple]{HM12} or directly measured in the simulations by \citet{Keating19} (brown), \citet{Cain21} (blue), \citet{Garaldi22} (orange; THESAN), and \citet{Lewis22} (red; CoDa-III).}
\label{fig:abcmfp2}
\end{figure*}

We apply the ABC inference methodology described above to the extended Ly$\alpha$ forest dataset from \citet{Bosman22}. The data consist of $51$ optical quasar spectra observed with the X-Shooter spectrograph \citep{Vernet11}, primarily sourced from the Extended XQR-30 sample \citep{D'Odorico23}, as well as $16$ spectra taken with the ESI \citep{Sheinis02} spectrograph, i.e. $67$ quasar spectra in total. All quasars are located at $z>5.7$ and their spectra possess signal-to-noise ratios (SNRs) larger than $10$ per pixel. Intervening damped Lyman-$\alpha$ absorbers are masked based on the detection of associated metal absorbers in the metal catalog of \citet{RDavies23}. In a few quasars, portions of the spectra with suspected broad absorption line contamination are also masked. We refer interested readers to \citep{Bosman22} for more details of the sample.

In order to measure the effective optical depths, the intrinsic quasar continua are reconstructed based on their un-absorbed emission features at $\lambda_{\text{rest}} > 1215.16$\AA. We use the principal component analysis (PCA) approach of \citet{Davies18a} with the improvements described in \citet{Bosman20b} and \citet{Bosman22}. The PCA approach reproduces the quasar continua at $\lambda_{\text{rest}}<1180$\AA \ with a wavelength-dependent $\sim8\%$ accuracy, and sub-percent precision. Finally, the effective optical depths are measured in bins of $dz = 0.1$ and only bins which are $>50\%$ un-masked are retained. The number of retained quasar sightlines $N_{\rm q}$ at each redshift is shown in Table~\ref{tab:results}. The masking, observational uncertainties, and wavelength-dependent continuum uncertainties are forward-modeled in all simulated mocks. \citet{Bosman22} showed that mock data generated in this manner is statistically indistinguishable from the real data in the post-reionization regime ($5.0<z<5.3$), solidifying our confidence in the noise model.

For each redshift bin -- the same $dz=0.1$ bins defined in \citet{Bosman22} -- we compute 10,000,000 mock data sets, drawing from uniform priors in $\Gamma_{\rm HI}$ and $\lambda_{\rm mfp}$ in uniform prior windows\footnote{In practice, we ran small numbers of mocks with very wide priors, and then progressively truncated them to optimize the sampling of the posterior core.} where the posterior PDF has substantial support. We then define a relative threshold $\epsilon_{\rm th}$ by the fraction of retained mocks with distances below the actual ABC threshold, e.g. with $\epsilon_{\rm th}=0.1$ we would retain the parameter values whose mock data sets resulted in a distance smaller than the 10th percentile of the entire distribution. For our fiducial posterior PDFs we retain the 1,000 samples with the smallest distances in our set of 10,000,000, i.e. $\epsilon_{\rm th}=10^{-4}$.

We show the resulting 2D posterior PDFs of $\Gamma_{\rm HI}$ and $\lambda_{\rm mfp}$ from $z=5.0$--$6.1$ in Figure~\ref{fig:abcpdfs}. The posteriors show only modest correlations between the two parameters, and both decrease steadily to higher redshift. While $\Gamma_{\rm HI}$ is well-constrained at all redshifts, along the $\lambda_{\rm mfp}$ dimension the posterior clearly runs into the edge of the prior (set by our grid of models) at $z\lesssim5.3$ and $z\gtrsim5.9$. To understand this apparent lack of constraining power, note that our constraints on $\lambda_{\rm mfp}$ are driven by a difference between the observed Ly$\alpha$ forest fluctuations and the fluctuations one would expect given a uniform ionizing background. At the low redshift end, the weak constraints are due to the fact that the observed Ly$\alpha$ forest fluctuations are fully consistent with a uniform ionizing background \citep{Bosman22}, while at the high-redshift end the strength of fluctuations is comparable to our shortest $\lambda_{\rm mfp}$ model. 

We note that the width of the posterior PDFs in the $\lambda_{\rm mfp}$ dimension are particularly large at $z=5.5$ and $z=5.9$ compared to their neighboring redshifts. In general, these variations come about due to the non-linear connection between $\lambda_{\rm mfp}$ and the width of the $\tau_{\rm eff}$ distribution -- the strength of radiation field fluctuations varies only modestly for $\lambda_{\rm mfp}\gtrsim40$\,Mpc (e.g. Figure~\ref{fig:uvbmap}), so any upward tail of the posterior will inevitably be elongated. In addition, at $z=5.5$ where this apparent difference is most pronounced, it was already noted in \citet{Bosman22} that the disagreement between the $\tau_{\rm eff}$ CDF and simulations assuming a uniform ionizing background was smaller than at any other redshift $z>5.3$. We can thus understand the relatively weak constraint, as this disagreement between the observations and uniform background simulations is the source of the constraining power of our analysis.

In Figure~\ref{fig:abcghi} we show the posterior medians (black circles) and central 68\% credible intervals (black error bars) of $\Gamma_{\rm HI}$, marginalized over $\lambda_{\rm mfp}$. We have also adjusted the posterior constraints by a few percent to correct for the coarse snapshot sampling employed in the simulated spectra (see Appendix~\ref{sec:appendix_snaps}); the ``raw'' uncorrected posterior medians are shown as open squares. Note that these statistical error bars apply only to the fiducial IGM thermal model with $z_{\rm re}=7.2$. The dark grey shaded regions in Figure~\ref{fig:abcghi} show the range of posterior medians obtained for different IGM thermal models with $z_{\rm re}=6.7$ and $z_{\rm re}=7.7$, while the light shaded region shows the range corresponding to extreme thermal models with $z_{\rm re}=6.2$ and $z_{\rm re}\sim15$ (see Section~\ref{sec:methods}). The uncertainty in $\Gamma_{\rm HI}$ resulting from the IGM thermal state is much larger than the statistical uncertainty except for $z\gtrsim5.9$, where the transmitted flux is much lower and sampled across relatively few sightlines. Tighter statistical constraints could likely be achieved at these redshifts from more informative summary statistics (e.g. the flux PDF on smaller scales and/or Ly$\beta$ transmission, cf. \citealt{Davies17}), but we leave a more detailed look at the $z\gtrsim6$ IGM in XQR-30 to future work.

The crosses in Figure~\ref{fig:abcghi} show the posterior medians in the self-consistent model (\S~\ref{sec:selfcon}), where the ionizing background fluctuations are drawn from the same physical location in the Nyx hydrodynamical box as the density field when computing the Ly$\alpha$ forest absorption. We see a gradual positive offset of the $\Gamma_{\rm HI}$ values in the self-consistent model which increases from a few percent at $z=5$ to $\sim25\%$ at $z\geq6$. This offset comes about due to correlation between ionizing background fluctuations and the density field -- regions with high $\Gamma_{\rm HI}$ tend to have higher density, and regions with low $\Gamma_{\rm HI}$ tend to have lower density, shifting the mean Ly$\alpha$ forest transmission to lower values. Thus the mean $\Gamma_{\rm HI}$ must be higher to reproduce the mean Ly$\alpha$ forest opacity. If we randomize the ionizing background fluctuations with respect to the density field in the $100$\,Mpc$/h$ box, we find that the resulting $\Gamma_{\rm HI}$ constraints are indistinguishable from the fiducial model.

In the left panel of Figure~\ref{fig:abcghi2}, we compare our constraints on $\Gamma_{\rm HI}$ (solid points) to literature values (open points) after combining the statistical uncertainties with the thermal state uncertainties in quadrature, along with an additional 0.03 dex of systematic uncertainty to approximate the uncertainty due to Jeans smoothing (e.g. \citealt{BB13}) and an additional upward systematic uncertainty given by the bias between the fiducial and self-consistent model constraints. We also show grey points which represent the posterior medians shifted upward by the same amount. Our constraints are consistent with previous measurements by \citet{Calverley11}, \citet{WB11}, and \citet{D'Aloisio18}, as well as constraints from the pilot study of \citet{Davies17}, although we note that we have not carefully accounted for the different IGM thermal states (or ranges of thermal states) assumed by those works. In the right panel of Figure~\ref{fig:abcghi2}, we compare to predictions of $\Gamma_{\rm HI}(z)$ from empirically-motivated 1D cosmological radiative transfer models by \citet{HM12}, \citet{KS19}, and \citet{FG20}, and the 3D coupled radiation-hydrodynamics simulations from \citet{Keating19}, THESAN \citep{Garaldi22}, and CoDa-III \citep{Lewis22}. While the 1D radiative transfer models are all in rough agreement with the average $\Gamma_{\rm HI}$ from $z\sim5$--$5.5$, they fail to reproduce the steep downturn towards $z\sim6$. In contrast, the $\Gamma_{\rm HI}(z)$ in the 3D simulations rises more rapidly, and is more consistent with the trend of our constraints. In particular, the evolution from the CoDa-III simulation closely reproduces the rapid rise from $z\sim6$ to $z\sim5.4$, although it overshoots to a more highly ionized IGM at $z\sim5$. 

In Figure~\ref{fig:abcmfp} we show the derived constraints on $\lambda_{\rm mfp}$. At $z=5.0$--$5.2$ and $z=6.0$--$6.1$, the posteriors show no clear peak interior to the boundaries of the prior. We define 95\% lower and upper limits, respectively, by the $\lambda_{\rm mfp}$ at which the posterior first falls a factor $e^{-2}$ from its peak value. At $z=5.3$ and $z=5.9$, the posterior PDFs are clearly peaked, but the posterior PDF does not quite fall below a factor of $e^{-2}$ from its peak value at the upper and lower edges of the prior boundary, respectively. We show these strongly prior-influenced constraints as open points with error bars. 

The open squares in Figure~\ref{fig:abcmfp} show the posterior medians for (random) ionizing background fluctuations drawn from the $100$\,Mpc$/h$ box. At redshifts where the mean free path is constrained to be quite small, $\lambda_{\rm mfp}\lesssim30$\,Mpc, this model prefers a shorter mean free path. This is due to the general decrease in the amplitude of background fluctuations in the smaller ionizing background simulation at fixed $\lambda_{\rm mfp}$. Quantitatively, we find that the width of the central 68\% distribution of fluctuations (in log $\tilde{\Gamma}$) is roughly a factor of two narrower in the $100$\,Mpc$/h$ box than the $512$\,Mpc box for $\lambda_{\rm mfp}=20$\,Mpc. For larger mean free paths, the situation can be reversed -- for $\lambda_{\rm mfp}\gtrsim50$\,Mpc, the larger-scale background fluctuations in the smaller volume exceed those in the larger one due to the truncation of the periodic boundary conditions in the \citet{DF16} method. 

The crosses in Figure~\ref{fig:abcmfp} show the posterior medians for the self-consistent model. In this case, we see a stark factor of $\sim2$ decrease in the preferred $\lambda_{\rm mfp}$ at all redshifts. This is due to the ionizing background fluctuations now having to overcome the large-scale fluctuations in the density field in order to increase the scatter in the Ly$\alpha$ forest opacity. That is, regions with a stronger ionizing background will preferentially lie in dense environments that will have a higher baseline Ly$\alpha$ opacity, while regions with weak ionizing background will correspond to voids. This then increases the requirements on the fluctuations in the radiation field relative to the uncorrelated case. However, the difference in $\lambda_{\rm mfp}$ is likely exaggerated by the relatively small volume of the Nyx simulation, $100$\,Mpc$/h$, compared to the fiducial $512$\,Mpc model, which lacks fluctuations in the source field on the largest coherence scales of the radiation field ($\gtrsim100$\,Mpc, see Figure~\ref{fig:uvbmap}), thus maintaining a stronger correlation between the background and the gas density on the smaller $dz=0.1\sim50$\,Mpc scale of the $\tau_{\rm eff}$ measurements. As we cannot assess the full strength of this effect, and how a much larger (and thus computationally very expensive) self-consistent model might mitigate this offset, we conservatively opt to extend the lower envelope of the uncertainty on $\lambda_{\rm mfp}$ by the magnitude of the measured offset between the self-consistent and fiducial method constraints from the 100\,Mpc$/h$ box. We also show the posterior medians shifted by this offset as grey points. At $z=5.5$ where the fiducial analysis using the 100\,Mpc$/h$ box results in a larger value for $\lambda_{\rm mfp}$ due to the additional effect of the truncated boundary conditions mentioned above, we instead adopt the difference between the self-consistent model and the 512\,Mpc model.

In the left panel of Figure~\ref{fig:abcmfp2}, we compare our constraints to the measurements from \citet{Worseck14}, \citet{Becker21}, and \citet{Zhu23}, as well as the lower limit from \citet{Bosman21MFP}. The single power-law evolution with redshift that \citet{Worseck14} find to be a good fit at $z=2$--$5$ is disfavored at $z\gtrsim5.4$, with our constraints indicating a more rapid decline to higher redshift. Intriguingly, our constraints are in good agreement with the $z=6$--$5$ trend found by \citet{Becker21} and \citet{Zhu23}, effectively bridging the gaps between the more direct quasar-stacking measurements. In the right panel of Figure~\ref{fig:abcmfp2}, we compare our constraints to theoretical predictions for the evolution of $\lambda_{\rm mfp}$. The empirical absorber model from \citet{HM12}, also adopted in a similar form by later ionizing background calculations (e.g.~\citealt{Puchwein19,FG20}), lies well above our constraints at $z>5.3$, suggesting that the distribution of \ion{H}{1} absorbers must evolve more rapidly at high redshift. Meanwhile, the sub-grid opacity model of the \citet{Cain21} simulations as well as the radiation-hydrodynamic simulations from \citet{Keating19}, THESAN \citep{Garaldi22}, and CoDa-III \citep{Lewis22} are in much better agreement, with perhaps a modest over-prediction of $\lambda_{\rm mfp}$ in the \citet{Keating19} simulations and THESAN relative to our constraints.

Our constraints on $\Gamma_{\rm HI}$ and $\lambda_{\rm mfp}$, including statistical-only and total uncertainties, are summarized in Table~\ref{tab:results}.

\begin{table*}
\centering
\begin{center}
\hskip -8em
\begin{tabular}{l c|c c c c c c c c}
$z$ & $N_{\rm q}$ & $\Gamma_{\rm HI}$ (10$^{-12}$\,s$^{-1}$) & stat.\,err. & tot.\,err & s-c\,corr. & $\lambda_{\rm mfp}$ (Mpc) & stat.\,err & tot.\,err & s-c\,corr. \\
\hline
 5.0 & 37 & 0.599 & $_{-0.031}^{+0.035}$ & $_{-0.085}^{+0.109}$ & 1.034 & -- & $>27.7$ & $>23.3$ & 0.828 \\
 5.1 & 48 & 0.582 & $_{-0.024}^{+0.028}$ & $_{-0.081}^{+0.115}$ & 1.043 & -- & $>33.7$ & $>26.1$ & 0.800  \\  
 5.2 & 55 & 0.547 & $_{-0.023}^{+0.022}$ & $_{-0.080}^{+0.104}$ & 1.035 & -- & $>42.6$ & $>33.9$ & 0.857  \\
 5.3 & 58 & 0.503 & $_{-0.021}^{+0.024}$ & $_{-0.074}^{+0.113}$ & 1.066 & 79.1 & $_{-29.7}^{+43.9}$ & $_{-52.3}^{+43.9}$ & 0.546 \\
 5.4 & 64 & 0.447 & $_{-0.024}^{+0.025}$ & $_{-0.069}^{+0.135}$ & 1.135 & 29.0 & $_{-5.2}^{+6.9}$ & $_{-17.5}^{+6.9}$ & 0.633 \\
 5.5 & 64 & 0.381 & $_{-0.017}^{+0.019}$ & $_{-0.060}^{+0.104}$ & 1.103 & 44.3 & $_{-11.3}^{+25.0}$ & $_{-31.7}^{+25.0}$ & 0.380 \\
 5.6 & 59 & 0.343 & $_{-0.021}^{+0.024}$ & $_{-0.055}^{+0.110}$ & 1.137 & 23.5 & $_{-4.7}^{+6.9}$ & $_{-13.9}^{+6.9}$ & 0.686 \\
 5.7 & 51 & 0.255 & $_{-0.016}^{+0.019}$ & $_{-0.042}^{+0.084}$ & 1.139 & 24.0 & $_{-6.2}^{+7.8}$ & $_{-15.1}^{+7.8}$ & 0.654 \\
 5.8 & 45 & 0.208 & $_{-0.019}^{+0.021}$ & $_{-0.036}^{+0.076}$ & 1.161 & 15.9 & $_{-4.8}^{+6.5}$ & $_{-10.1}^{+6.5}$ & 0.652 \\
 5.9 & 28 & 0.185 & $_{-0.020}^{+0.021}$ & $_{-0.035}^{+0.063}$ & 1.125 & 23.5 & $_{-10.6}^{+19.0}$ & $_{-17.8}^{+19.0}$ & 0.474 \\
 6.0 & 19 & 0.167 & $_{-0.029}^{+0.032}$ & $_{-0.037}^{+0.079}$ & 1.195 & -- & $<22.6$ & $<22.6$ & --  \\
 6.1 & 10 & 0.147 & $_{-0.039}^{+0.055}$ & $_{-0.044}^{+0.097}$ & 1.229 & -- & $<21.6$ & $<22.6$ & --  \\
 \hline
\end{tabular}
\end{center}
\caption{Derived constraints on the hydrogen photoionization rate $\Gamma_{\rm HI}$ in units of $10^{-12}$\,s$^{-1}$ and mean free path $\lambda_{\rm mfp}$ in units of comoving Mpc. The nominal values represent the median of the posterior PDF, with the statistical error (stat. err.) representing the 16th to 84th percentiles of the posterior PDF. The total error (tot. err.) includes contributions from systematic uncertainties (see text for details), and should be considered highly covariant between redshift bins. The self-consistency correction (s-c corr.) is the ratio of the posterior medians of the self-consistent model and the fiducial model; the grey points in Figure~\ref{fig:abcghi2} and Figure~\ref{fig:abcmfp2} can be recovered by multiplying the fiducial constraints by the value in this column. Upper and lower limits on $\lambda_{\rm mfp}$ roughly correspond to $2\sigma$ limits, see \S~\ref{sec:constraints} for details.}
\label{tab:results}
\end{table*}

\subsection{Consistency between data and model}

As in most other Bayesian parameter inference methods, our ability to place constraints on model parameters does not require that the model actually provides a good description of the data. Here we investigate the degree to which the data are consistent with being a draw from our model.

\begin{figure*}
\begin{center}
\resizebox{18cm}{!}{\includegraphics[trim={0em 1em 1em 1em},clip]{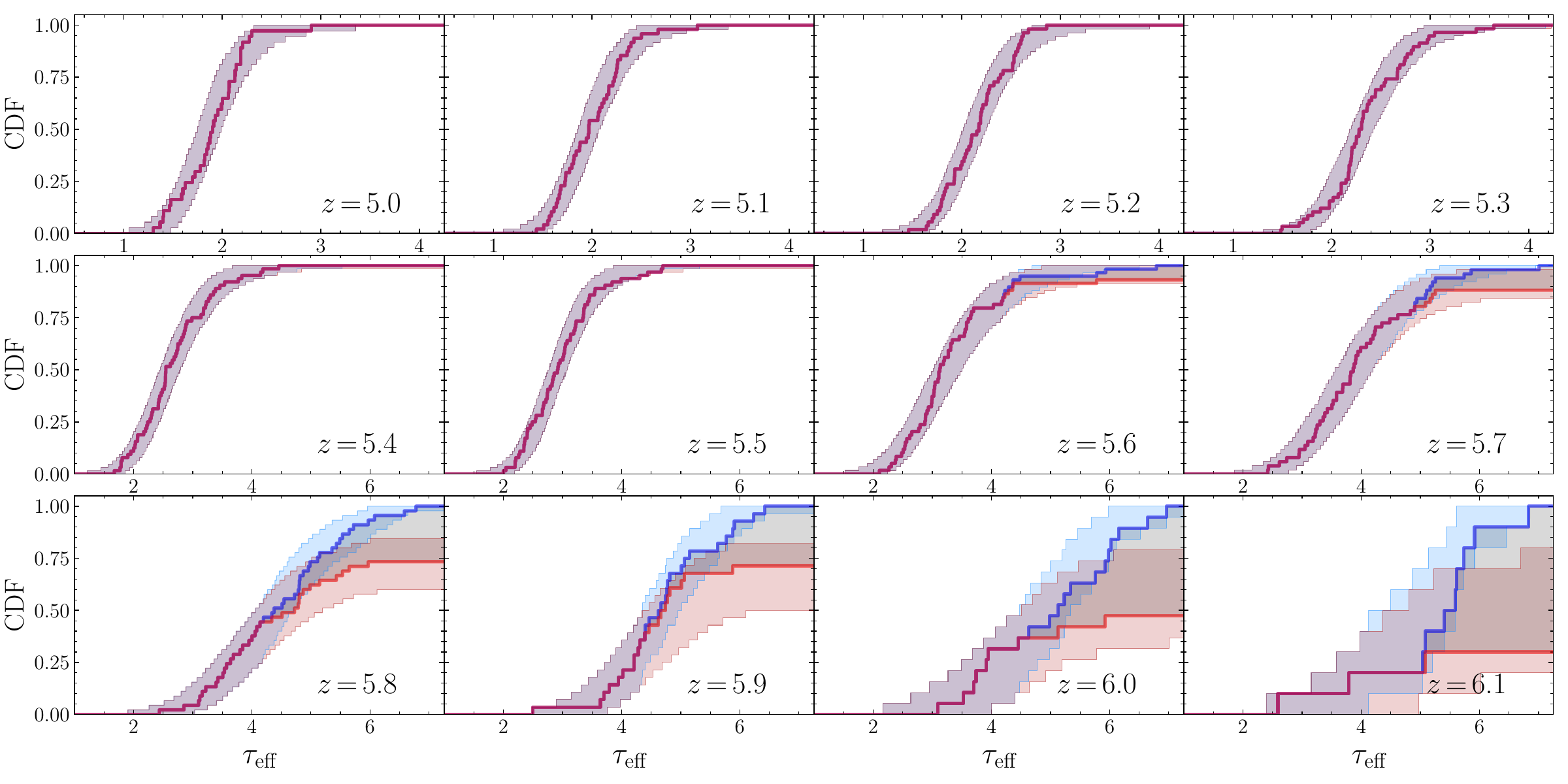}}
\end{center}
\caption{Comparison between the Ly$\alpha$ forest $\tau_{\rm eff}$ CDFs from \citet{Bosman22} (solid curves) and the central 95\% of the distribution of mock CDFs from our simulations when adopting the posterior mean values of $\Gamma_{\rm HI}$ and $\lambda_{\rm mfp}$ (shaded regions). The red color corresponds to CDFs where Ly$\alpha$ fluxes detected at less than $2\sigma$ significance are assumed to be zero (i.e. $\tau_{\rm eff}=\infty$), while the blue color corresponds to those fluxes being set to twice the statistical noise (i.e., $\tau_{\rm eff}=-\ln(2\times\sigma_{N})$).}
\label{fig:cdfcompare}
\end{figure*}

In Figure~\ref{fig:cdfcompare}, we show the CDFs of the observed Ly$\alpha$ forest data from \citet{Bosman22} as blue and red solid lines, corresponding to the ``optimistic'' and ``pessimistic'' definitions from \citet{Bosman18}, similar to Figure~\ref{fig:abcexample}. In the former case, Ly$\alpha$ mean fluxes below twice the (statistical) noise $\sigma_F$ are set to $\tau_{\rm eff}=-\ln{(2\times\sigma_F)}$, while in the latter case, they are assumed to have $\tau_{\rm eff} > 10$. The value of the red curves at the right-hand edge of the panels in Figure~\ref{fig:cdfcompare} thus corresponds to the fraction of Ly$\alpha$ forest sightlines with detected (i.e. $>2\sigma_F$) flux. The shaded regions correspond to the central 95\% of the distribution of mock data sets drawn from the posterior mean values of $\Gamma_{\rm HI}$ and $\lambda_{\rm mfp}$, where blue and red similarly correspond to the optimistic and pessimistic CDF definitions. We can see that in the majority of cases, the red and blue curves fit neatly within the model distributions, with the only exceptions being a handful of the most highly opaque sightlines at $z=5.6$ and $z=5.8$. In the highest redshift bins, $z=6.0$ and $z=6.1$, the blue curves lie close to the upper end of the model distributions -- however, this is the expected behavior in the regime where most observations result in non-detections, as the ``optimistic'' CDF has an upper bound in $\tau_{\rm eff}$ given by the distribution of noise values of the quasar spectra, i.e. the CDF of $-\ln{(2\times\sigma_F)}$.

\begin{figure*}
\begin{center}
\resizebox{18cm}{!}{\includegraphics[trim={0em 1em 1em 1em},clip]{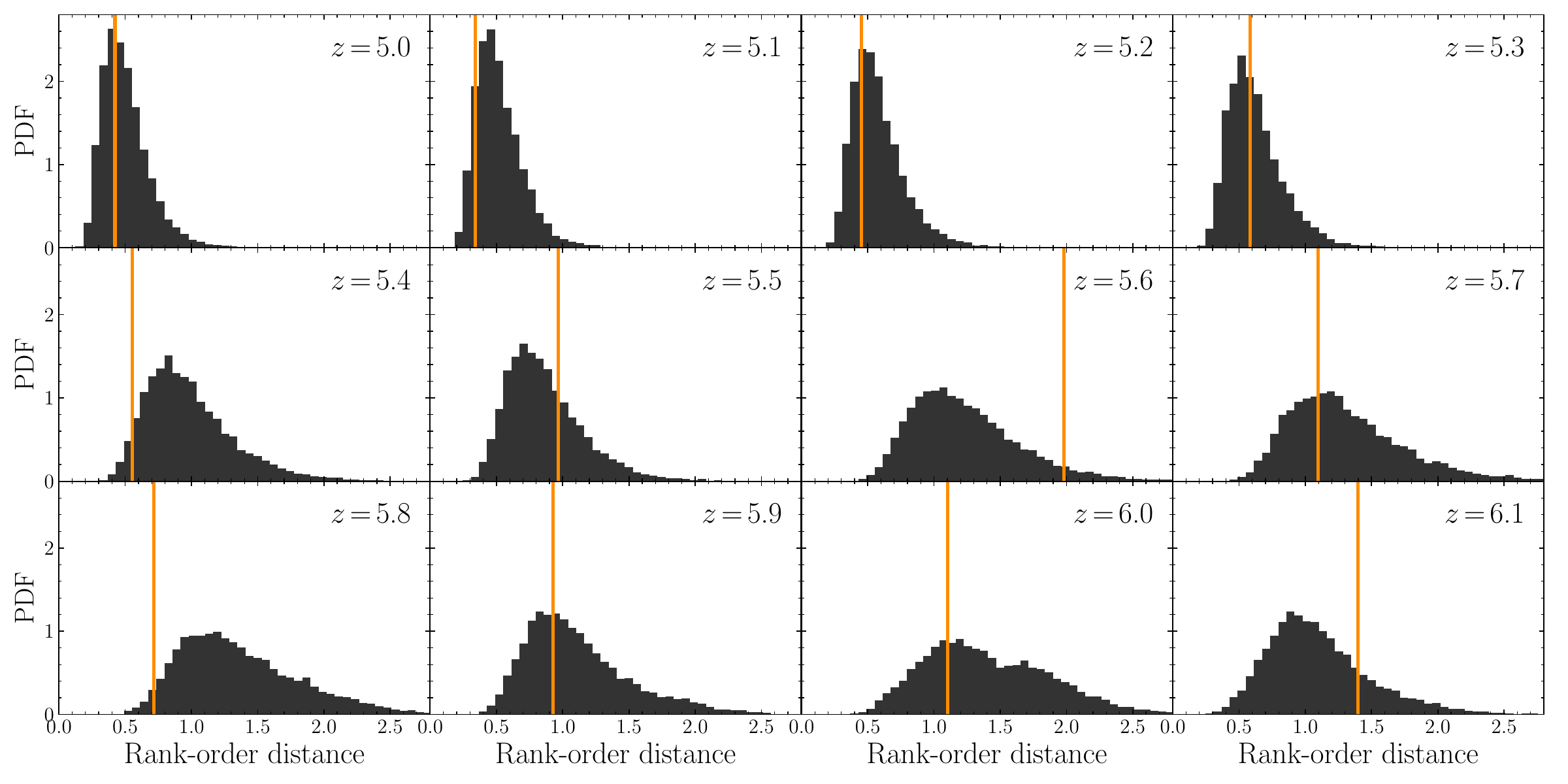}}
\end{center}
\caption{Distribution of distances (equation~\ref{eqn:abcdist}) between the average set of rank-order $\teff$ and individual sets from 10,000 mock observations, both adopting the posterior mean values of $\Gamma_{\rm HI}$ and $\lambda_{\rm mfp}$ at each redshift. The vertical orange lines show the distance between the average simulated rank-order $\teff$ and the observed Ly$\alpha$ forest data.}
\label{fig:consistency}
\end{figure*}

To further quantify the goodness-of-fit, we compare the ABC distance between the average simulated CDF at the posterior mean values of $\Gamma_{\rm HI}$ and $\lambda_{\rm mfp}$ and the observed data to the distribution of distances to 10,000 mock data sets generated from the same model. We show the distribution of the resulting consistency metric in Figure~\ref{fig:consistency}. In general, the distance to the data is well within the range of distances to mock data sets.

\section{Caveats and Limitations of our Analysis}\label{sec:caveatemptor}

As mentioned previously, our fiducial model for post-reionization ionizing background fluctuations suffers from important limitations. In this section, we reiterate and summarize these limitations, and discuss their potential consequences for the interpretation of our results.

\subsection{Uncorrelated density and ionization rate fields}

In the fiducial model, by virtue of using separate volumes for the ionizing background calculation and the hydrodynamical simulation, the density and ionizing radiation fields of the mock spectra used for inference are decoupled. The two should generally be correlated on large scales (e.g. \citealt{MF09}). Because the Ly$\alpha$ forest opacity roughly scales as $\tau_{\rm Ly\alpha}\propto \Delta^2 \Gamma_{\rm HI}^{-1}$ (e.g. \citealt{Weinberg97}), the variance of $\tau_{\rm Ly\alpha}$ should behave as $\sigma_{\tau}^2\sim\tau^2[\sigma_{\Delta^2}^2/\Delta^2+\sigma_{\Gamma}^2/\Gamma-2\sigma_{\Delta^2 \Gamma}/\Delta^2\Gamma]$, i.e. the presence of correlated fluctuations in the density and ionization rate ($\sigma_{\Delta^2 \Gamma}$) should suppress the strength of Ly$\alpha$ forest fluctuations at fixed $\lambda_{\rm mfp}$. Because the fluctuations become stronger with decreasing $\lambda_{\rm mfp}$, we expect our constraints to be biased high.

We tested this hypothesis by constructing a self-consistent model of ionizing background fluctuations within the smaller 100\,Mpc$/h$ volume of our hydrodynamical simulation (\S~\ref{sec:selfcon}). As expected, in the regime of strong Ly$\alpha$ forest fluctuations we recover much shorter mean free paths. A fraction of this difference is due to the suppression of the amplitude of fluctuations by the relatively small volume of the Nyx simulation, $100$\,Mpc$/h$ on a side, compared to the fiducial ionizing background model, $512$\,Mpc on a side. Without a much larger self-consistent model we are unable to entirely disentangle these two effects, so we instead opt to conservatively allow for a systematic error that encompasses the constraints derived from the self-consistent model, as seen by the large lower error bars in Figure~\ref{fig:abcmfp2}.

\subsection{Uncertainties in the IGM thermal state}\label{sec:thermal}

As previously mentioned in \S~\ref{sec:methods}, our hydrodynamical simulation was run with an optically thin UV background which reionized and heated the volume at very early times, $z_{\rm re}\sim15$. We have chosen to adjust the temperatures in post-processing to better reflect current constraints on the timing and heat injection of reionization. In particular, we assume a fixed heat injection of $\Delta T=20,000$\,K, which is a representative temperature of the gas after the passage of the ionization front \citep{MR94,D'Aloisio18b}, and tuned the range of $z_{\rm re}$ to be consistent with the IGM thermal state measured from the distribution of Ly$\alpha$ transmission spike widths by \citet{Gaikwad20} (see also \citealt{Bolton12}). In principle, one could treat $\Delta T$ and $z_{\rm re}$ as additional parameters, introduce priors, and add the \citet{Gaikwad20} measurements to the computation of the likelihood. Due to the computational demands of our likelihood-free inference method, and because in this work we are not attempting to constrain $\Delta T$ or $z_{\rm re}$, we opted to instead impose a plausible range rather than include them directly in the parameter inference.

We note, however, that the IGM thermal state at $z>5$ is still substantially uncertain. Measurements using the 1D Ly$\alpha$ forest flux power spectrum by \citet{Walther19} and \citet{Boera19} suggest somewhat lower temperatures with a steeper temperature-density relation at $z\sim5$. If the IGM was reionized earlier, or if the heat injection was much lower, the upper envelope of the light grey region in Figure~\ref{fig:abcghi} shows that the $\Gamma_{\rm HI}$ required to reproduce the observed Ly$\alpha$ forest transmission could be substantially higher, with less evolution required from $z\sim6$ to $z\sim5$.

While our hydrodynamical simulation includes the effect of Jeans smoothing on the gas via its fiducial thermal history, by post hoc altering the IGM thermal state of the simulation we neglect the differences in this smoothing that the different thermal histories would have otherwise imprinted (e.g. \citealt{GH98,Peeples10a,Peeples10b,Kulkarni15,Nasir16,Onorbe17}). The effect of Jeans smoothing on measurements of $\Gamma_{\rm HI}$ was thoroughly investigated over a wide range of thermal histories by \citet{BB13}, who found that at $z\lesssim5$ the effect was minor, contributing $\lesssim0.03$ dex to the error budget. We adopt a somewhat higher $0.03$ dex systematic uncertainty to account for the trend of larger error at higher redshift, and add this additional error (in quadrature) to our fiducial constraints.

\subsection{Lack of post-reionization temperature fluctuations}

The thermal state of our hydrodynamical simulation assumes a homogeneous heat injection, but the reionization process is patchy with different reionization timing in different locations \citep{Furlanetto04}. This should lead to a highly inhomogeneous IGM thermal state immediately after reionization is complete, which persists to much later times (e.g. \citealt{Trac08,Onorbe18,Wu19,Keating18,D'Aloisio18b}). The thermal state in our simulation is thus too uniform, and lacks any {(anti-)correlation} with large-scale density. Due to the expected anti-correlation between large-scale density and post-reionization temperature fluctuations \citep{D'Aloisio15}, our Ly$\alpha$ forest fluctuations are likely overestimated at fixed $\lambda_{\rm mfp}$ \citep{Davies17b,D'Aloisio18}. This would act to bias our $\lambda_{\rm mfp}$ measurements high; with a more realistic simulation, we would require stronger ionizing background fluctuations to counteract the effect of thermal state fluctuations and reproduce the large variations in the Ly$\alpha$ forest, and thus estimate a lower $\lambda_{\rm mfp}$.

\subsection{Parameter choices in the fluctuating ionizing background simulations}

Our model for ionizing background fluctuations has several fixed parameters that we have not explored in detail. For example, we assume that $\lambda_{\rm mfp}\propto\Delta^{-1}\Gamma_{\rm HI}^{2/3}$, but both power-law indices are uncertain. \citet{Chardin17} found a shallower density dependence $\lambda_{\rm mfp}\propto\Delta^{-0.4}$ when post-processing the Sherwood simulations \citep{Bolton17} with the self-shielding prescription from \citet{Rahmati13}. Incorporating this weaker density dependence would lead to stronger ionizing background fluctuations at fixed $\lambda_{\rm mfp}$. Our choice of the $\Gamma_{\rm HI}$ dependence is motivated by \citet{McQuinn11}, but both higher (up to $\sim1$) and lower values (down to $\sim1/3$) are plausible (see the discussion in \citealt{Becker21}), which would increase or decrease the $\lambda_{\rm mfp}$ required to match the observed Ly$\alpha$ forest fluctuations, respectively. 

In addition, our model for the ionizing sources assumes that only halos more massive than $2\times10^9$\,$M_\odot$ produce ionizing photons, and that these halos can be assigned a UV luminosity via abundance matching, and further that the ionizing luminosity is proportional to the UV luminosity. All of these assumptions impact the effective bias of the ionizing emissivity field in the fluctuating ionizing background simulations. Other works have adopted different combinations of these assumptions -- for example, \citet{Kulkarni19} prescribe ionizing luminosities proportional to halo mass for $M_h>10^9$\,$M_\odot$, which results in more ionizing photons coming from lower mass halos and thus a lower bias of the emissivity field relative to our approach. The possible parameter space of source models is quite large, requiring far more efficient approaches to statistical inference than ours to constrain these additional parameters (e.g. \citealt{Qin21}). 

\section{Discussion}

With the caveats above in mind, the constraints presented in this work show a substantial improvement in the statistical precision of $\Gamma_{\rm HI}(z)$ at $z>5$, and provide the first quantitative estimates of $\lambda_{\rm mfp}$ from the excess fluctuations in the Ly$\alpha$ forest alone. In this section we discuss the implications of these measurements for our understanding of the $z>5$ IGM.

\subsection{Is late reionization required by $z\sim5$--$6$ Ly$\alpha$ forest fluctuations?}

Since the successful reproduction of the large-scale $z\sim5$--$6$ Ly$\alpha$ forest variations by \citet{Kulkarni19}, it has commonly been understood that reionization is incomplete at least down to $z\sim5.5$ (and more recently, $z\sim5.3$, cf. \citealt{Bosman22,Zhu21,Zhu22}). However, here we have demonstrated good agreement between the data and a model which does not require incomplete reionization, which at first glance goes against this consensus. Crucially, due to the lack of full self-consistency in our model described in Section~\ref{sec:methods}, we cannot make a strong claim that reionization must be complete at $z\lesssim6$. That said, previous claims of a complete incompatibility between the Ly$\alpha$ forest opacity distribution at $z\sim5.5$--$6$ and a fully ionized IGM may not be entirely conclusive.

The IGM models employed by works in the literature that have suggested that incomplete reionization is required at $z\sim5.5$--$6$ are not without their own limitations. The moment-based radiative transfer method used by \citet{Kulkarni19} and \citet{Keating19} has been suggested to exhibit suppressed fluctuations in the radiation field at the end of reionization \citep[][see also \citealt{Gaikwad23}]{Wu21}, thus they may require more large-scale neutral islands from incomplete reionization in order to achieve strong Ly$\alpha$ forest fluctuations. In addition, the spatial resolution of their radiative transfer models ($\sim100$\,kpc) is too coarse to resolve self-shielding in Lyman limit systems (e.g. \citealt{McQuinn11}), potentially resulting in an artificially elongated mean free path in ionized regions and further reducing the strength of fluctuations. The semi-numerical method of \citet{Qin21} similarly requires a substantial neutral fraction ($x_{\rm HI}\sim0.15$) to reproduce the Ly$\alpha$ forest transmission statistics at $z\sim5.8$ from \citet{Bosman18}, but they employed an approximate spatial filtering approach to compute ionizing background fluctuations. In contrast, the semi-numerical method of \citet{Choudhury21} employed a somewhat more sophisticated (but similar) treatment of the radiation field, and required roughly half as much neutral gas to explain the same Ly$\alpha$ forest data set. Similar to our findings, \citet{Zhu21,Zhu22} found that the early reionization/short mean free path model from \citet{ND20} is consistent with the distribution of dark gaps in the Ly$\alpha$ and Ly$\beta$ forests.

Finally, we emphasize that our results do not in any way rule out the presence of a significant neutral hydrogen fraction in the $z<6$ IGM. Rather, they show that imposing the existence of neutral islands is not explicitly required to match the particular summary statistic we are considering, namely the distribution of $\tau_{\rm eff}$ on $dz=0.1$ scales. In reality, the $\lambda_{\rm mfp}$ we measure is only representative of the ``actual'' mean free path if the universe is fully ionized as assumed in our simulations. Our $\lambda_{\rm mfp}$ may thus be better interpreted as an effective parameter, reflecting both the existence of neutral islands in the deepest, large-scale voids, but also the strong fluctuations in the ionizing background due to a short, fluctuating mean free path of ionizing photons inside of ionized regions. Indeed, in our shortest mean free path models, large-scale regions with the weakest ionizing background are consistent with the gas remaining mostly neutral. However, compared to \citet{Gaikwad23} who see a similar effect, we estimate lower volume-averaged neutral fractions of $\sim0.02$--$7\%$ at $z=5.8$ ($\sim0.02$--$8\%$ at $z=5.9$), with a large systematic uncertainty between our fiducial and self-consistent models. We will explore the nature of these regions more closely in future work.

\subsection{Constraints on the ionizing emissivity}

\begin{figure}
\begin{center}
\resizebox{8cm}{!}{\includegraphics[trim={1.0em 1em 1.0em 1em},clip]{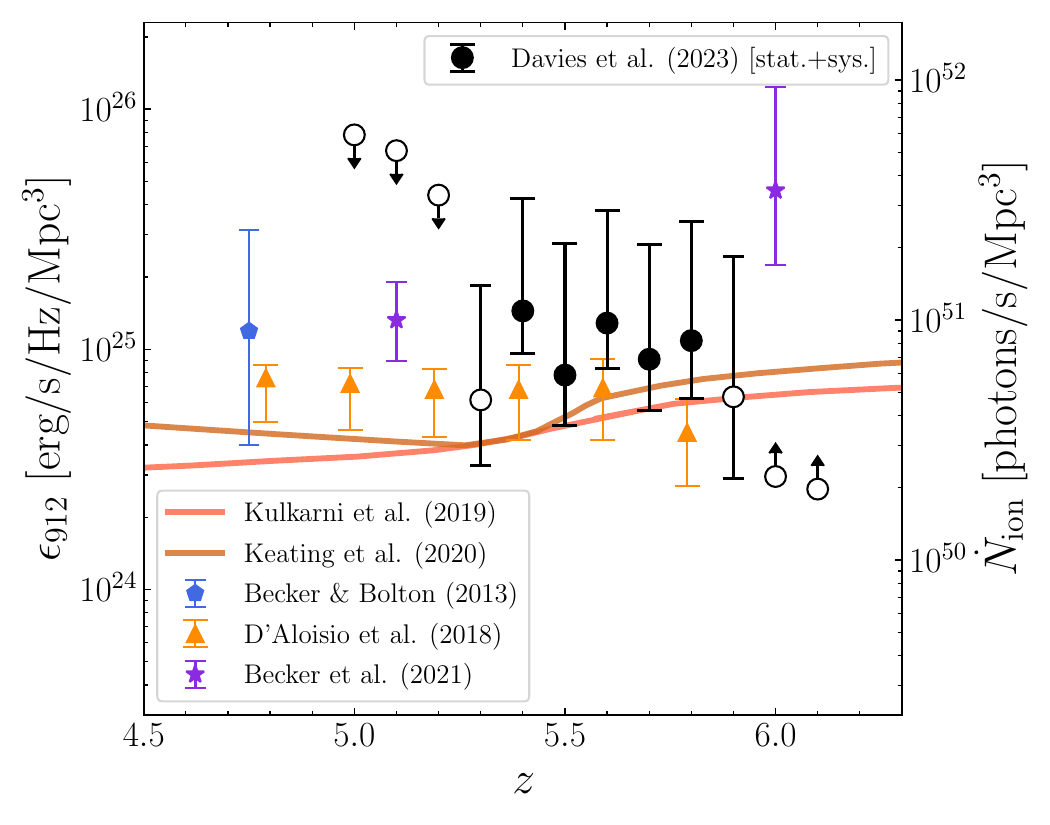}}
\end{center}
\caption{Implied ionizing emissivity $\epsilon_{912}$ evolution from our $\Gamma_{\rm HI}$ and $\lambda_{\rm mfp}$ constraints (black circles) compared to literature values from \citet{BB13} (blue pentagon), \citet{D'Aloisio18} (orange triangles), and \citet{Becker21} (purple stars). The right axis shows the corresponding photon emissivity $\dot{N}_{\rm ion}$ assuming an ionizing source spectral index $\alpha=2$.}
\label{fig:emiss}
\end{figure}

\begin{figure*}
\begin{center}
\resizebox{8.5cm}{!}{\includegraphics[trim={1.0em 1em 1.0em 1em},clip]{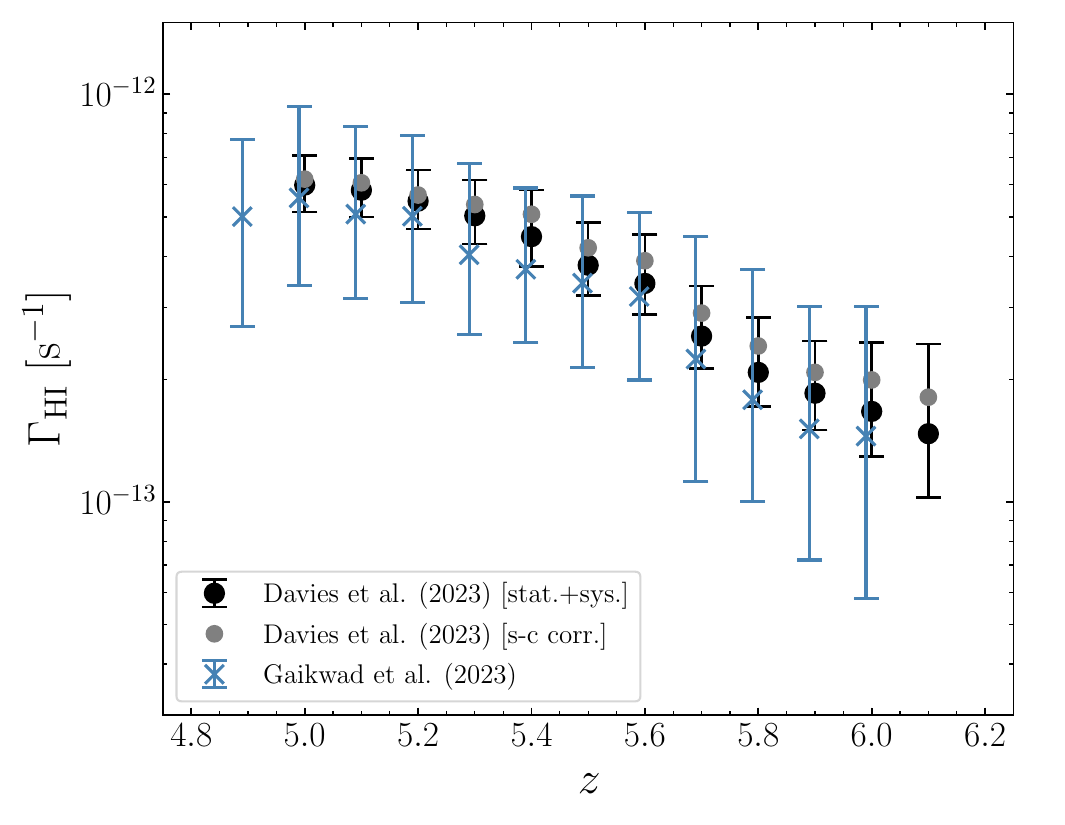}}
\resizebox{8.5cm}{!}{\includegraphics[trim={1.0em 1em 1.0em 1em},clip]{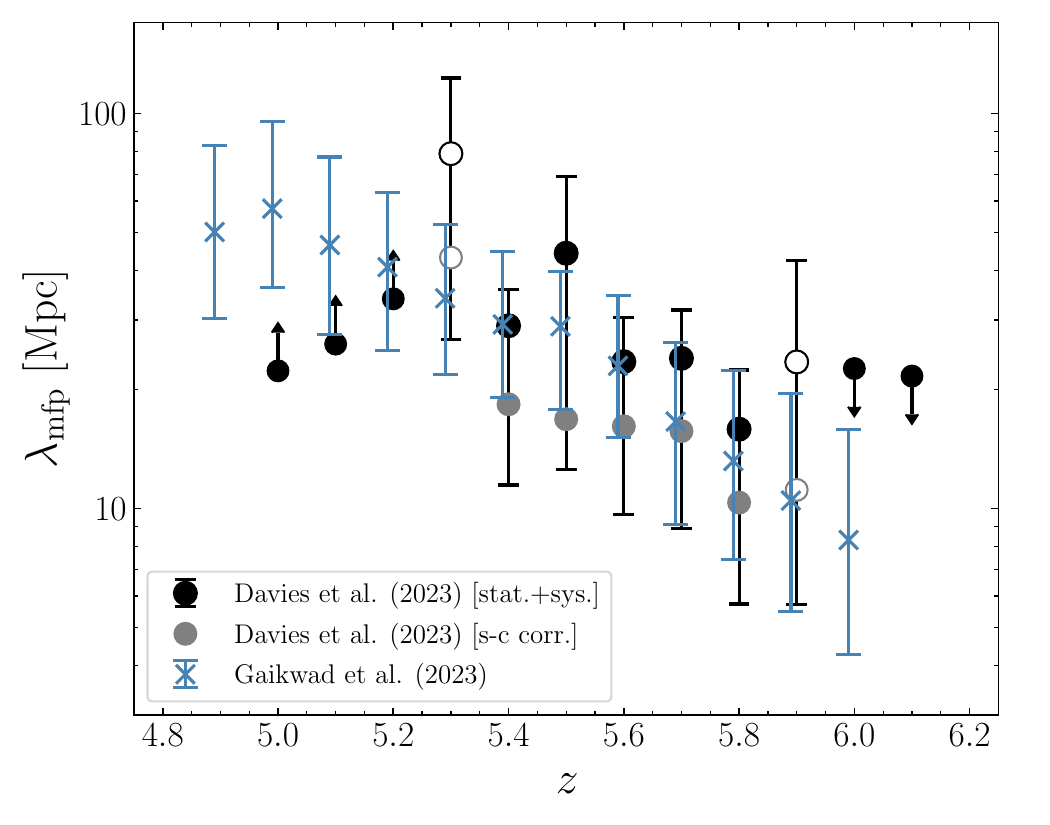}}
\end{center}
\caption{Left: Our constraints on $\Gamma_{\rm HI}$ (black and grey circles) compared to \citet{Gaikwad23} (blue crosses), where we have shifted the latter slightly in redshift for clarity. The mean trend agrees very well, but the wider assumed range in IGM thermal state parameters leads to larger error bars in \citet{Gaikwad23}. Right: Comparison to the mean free path constraints from \citet{Gaikwad23}. Despite the substantial philosophical differences in the definition of $\lambda_{\rm mfp}$ (see text for details), our constraints agree quite well.}
\label{fig:gaikwad}
\end{figure*}

While the mean free path we measure is highly uncertain, and may be more of an effective parameter as described above, we can nevertheless cautiously explore an interpretation of our $\Gamma_{\rm HI}$ and $\lambda_{\rm mfp}$ constraints in terms of the ionizing emissivity at the hydrogen-ionizing edge, $\epsilon_{912} = \epsilon_\nu(\lambda=912\,{\rm \AA})$. Assuming the local source approximation \citep{MW03}, the relationship between $\Gamma_{\rm HI}$, $\lambda$, and $\epsilon$ can be expressed as
\begin{eqnarray}
    \Gamma_{\rm HI} &=& 4\pi \int_{\nu_{\rm HI}}^{\infty} \frac{J_\nu}{h\nu}\sigma_{\rm HI}(\nu)d\nu \nonumber\\ 
    &\approx& \int_{\nu_{\rm HI}}^{\infty} \frac{\epsilon_\nu \lambda_\nu}{h\nu}\sigma_{\rm HI}(\nu)d\nu.
\end{eqnarray}
In practice, at $z>5$ the integral over frequency effectively extends only to $4\times\nu_{\rm HI}$ due to the onset of strong \ion{He}{2} absorption. Approximating the ionizing emissivity as a power law $\epsilon_\nu = \epsilon_{912} (\nu/\nu_{\rm HI})^{-\alpha}$, and recalling that we also treat the mean free path as a power law with frequency (equation~\ref{eqn:mfplaw}), we numerically solve this expression for $\epsilon_{912}$ at each redshift using our measured $\Gamma_{\rm HI}$ and $\lambda_{\rm mfp}$ values. We assume a fiducial emissivity spectral index of $\alpha=2$, following \citet{BB13}. 

Our ionizing emissivity estimates are subject to the statistical and systematic uncertainties in the $\Gamma_{\rm HI}$ and $\lambda_{\rm mfp}$ measurements described above, as well as an additional systematic uncertainty in the spectral index $\alpha$ assumed for the ionizing sources. Following \citet{BB13}, we adopt a range of $\alpha=1.0$--$3.0$. For our fiducial uncertainty estimates shown in Figure~\ref{fig:emiss}, we propagate the statistical uncertainties assuming $\epsilon_{912}\propto\Gamma_{\rm HI}/\lambda_{\rm mfp}$, and combine them in quadrature with the systematic uncertainties on $\Gamma_{\rm HI}$ and $\lambda_{\rm mfp}$ as well as the systematic uncertainty from the spectral index variations $\alpha=1.0$--$3.0$. For simplicity, here we treat the self-consistent model corrections for $\Gamma_{\rm HI}$ and $\lambda_{\rm mfp}$ by inflating the systematic error terms in the upper and lower directions, respectively.

The resulting ionizing emissivities and their corresponding uncertainties are shown in Figure~\ref{fig:emiss}, compared to literature constraints from \citet{BB13}, \citet{D'Aloisio18}, and \citet{Becker21}. Note that $\epsilon_{912}$ is related to the number of ionizing photons emitted per comoving volume $\dot{N}_{\rm ion}$ by $\dot{N}_{\rm ion} = h\alpha\epsilon_{912}$ (e.g. \citealt{BB13}), shown on the right axis assuming $\alpha=2$. We find that the ionizing emissivity is consistent with a roughly constant $\epsilon_{912}\sim10^{25}$\,erg\,s$^{-1}$\,Hz$^{-1}$\,Mpc$^{-3}$ from $z\sim6$--$5$, consistent with previous work.

\subsection{Comparison to XQR-30 analysis by Gaikwad et al.}

In a recent complementary work, \citet{Gaikwad23} have also constrained the photoionization rate and ionizing photon mean free path from the same XQR-30 Ly$\alpha$ forest data that we employ here. Here we compare our results and discuss the differences in analysis methodology.

\citet{Gaikwad23} use a self-consistent model (in our terminology, cf. \S~\ref{sec:selfcon}) consisting of smoothed-particle-hydrodynamics (SPH) simulations with $2048^3$ baryon and dark matter particles in a volume $160$\,Mpc$/h$ on a side. They compute ionizing background fluctuations using an independently developed version of the \citet{DF16} method which is heavily optimized by using a tree decomposition of the emissivity field, allowing them to evaluate the radiation field at a much higher resolution ($512^3$, $\sim0.5$\,Mpc). The statistical comparison between their dense model grid and the XQR-30 Ly$\alpha$ forest data is performed using the Anderson-Darling test on the $\tau_{\rm eff}$ CDF, where they set a p-value threshold to map out a $1-\sigma$ contour around the best-fit model. 

The largest difference between the two works is primarily a philosophical one -- \citet{Gaikwad23} perform inference on a different definition of the mean free path $\lambda_{\rm mfp}$ than adopted here. Specifically, while we treat the mean free path as a fully sub-grid quantity that arises from unresolved gas clumping, \citet{Gaikwad23} measure $\lambda_{\rm mfp}$ from the density field of their simulation after applying the fluctuating $\Gamma_{\rm HI}$ field. That is, for a large number of skewers, they evaluate the density of neutral hydrogen $n_{\rm HI}$ given the local gas density and $\Gamma_{\rm HI}$, and then calculate the ionizing opacity at the Lyman limit $\tau_{\rm HI}$ by integrating the contribution from each resolution element,
\begin{equation}
    \tau_{\rm HI}(x) = \int_0^x n_{\rm HI} \sigma_{\rm HI}(\nu_{\rm HI})\,dl.
\end{equation}
They then compute the corresponding mean free path by fitting an exponential profile to the average transmission ($F(x)=\exp{[-\tau_{\rm HI}(x)]}=\exp{[-x/\lambda_{\rm mfp}]}$) of the simulated skewers. This procedure imposes a physical prior on the possible combinations of $\Gamma_{\rm HI}$ and $\lambda_{\rm mfp}$ via the gas density distribution of their hydrodynamical simulation. In addition, this allows \citet{Gaikwad23} to obtain tight constraints on $\lambda_{\rm mfp}$ even without an excess in Ly$\alpha$ forest fluctuations, as it can be obtained from the density field even if the ionizing background is entirely uniform. In contrast, our method infers the mean free path solely from the excess in Ly$\alpha$ forest fluctuations over the uniform case, i.e. from the character of the radiation field fluctuations alone.

In Figure~\ref{fig:gaikwad}, we compare our constraints on $\Gamma_{\rm HI}$ (left) and $\lambda_{\rm mfp}$ (right) to \citet{Gaikwad23}. In general the two agree very well, suggesting that despite our rather different statistical and modeling methodologies, the rapid evolution in both $\Gamma_{\rm HI}$ and $\lambda_{\rm mfp}$ is robustly indicated by the XQR-30 Ly$\alpha$ forest data. In detail, our method recovers a smaller uncertainty on $\Gamma_{\rm HI}$, primarily due to the broader range of IGM thermal parameters marginalized over by \citet{Gaikwad23}, while they achieve tighter constraints on $\lambda_{\rm mfp}$ more uniformly across the full redshift range of their study, subject to the difference in $\lambda_{\rm mfp}$ definition described above.

\section{Conclusion}

In this work, we have constrained the evolution of the photoionization rate $\Gamma_{\rm HI}$ and the mean free path of ionizing photons $\lambda_{\rm mfp}$ in the IGM at $z=5.0$--$6.1$ using the extended XQR-30 \citep{D'Odorico23} Ly$\alpha$ forest data set from \citet{Bosman22}. We assume that the excess fluctuations in the $z>5.3$ Ly$\alpha$ forest are due to a strong coupling between the ionizing background and mean free path \citep{DF16}, and constrain parameters using the likelihood-free inference technique of approximate Bayesian computation, or ABC. 

We recover a smooth evolutionary trend in $\Gamma_{\rm HI}$, which increases by a factor of $\sim4$ from $z=6$ to $z=5$; in agreement with past observations but with a finer redshift sampling and smaller statistical uncertainty. The increase is much stronger than predicted by empirically-motivated 1D cosmological radiative transfer models, but is largely consistent with the evolution found by state-of-the-art 3D radiation-hydrodynamic cosmological simulations. We similarly find consistency with recent measurements of the mean free path $\lambda_{\rm mfp}$ from stacked quasar spectra and recent radiation-hydrodynamic models. 

We find that the statistical constraining power of the coarsely-binned $\tau_{\rm eff}$ measurements is very strong, to the extent that for both $\Gamma_{\rm HI}$ and $\lambda_{\rm mfp}$ we are strongly limited by systematic uncertainties. For the former, we are limited by our knowledge of the IGM thermal state, while for the latter (and the former, to a lesser extent) we are limited by the computational resources required to simulate a converged Ly$\alpha$ forest in a large enough volume to capture the full intensity of ionizing background fluctuations. Additionally, the relic thermal fluctuations left by reionization at earlier times (e.g. \citealt{D'Aloisio15}) will also confound efforts to precisely constrain the parameters of the $z=5$--$6$ IGM without a complete model of the entire reionization process. Reducing these systematic uncertainties will require extracting information from both the small-scale (e.g. \citealt{Gaikwad20}) and large-scale properties of the Ly$\alpha$ forest, and then performing statistical inference on full reionization lightcones with a detailed model for the sources and sinks (e.g. \citealt{Qin21}). 

At the low redshift end of our data set, $z\sim5$, our ability to constrain $\lambda_{\rm mfp}$ is limited by the insensitivity of the Ly$\alpha$ forest to long mean free paths, as the resulting fluctuations in the radiation field become weak compared to the intrinsic fluctuations already imprinted by the IGM density field. However at these redshifts the direct stacking of quasar spectra appears to be a far more sensitive probe with fewer underlying assumptions and systematic uncertainties \citep{Worseck14,Becker21,Zhu23}. At the high redshift end, $z\sim6$, we are instead limited by the relative sparsity of sightlines and the poor spatial resolution of our ionizing background simulations. The latter can already be rectified by adopting more efficient methods of computing the radiation field (e.g. \citealt{Gaikwad23}), while the former will require additional deep spectroscopy of yet higher redshift quasars than the XQR-30 sample (e.g. \citealt{Yang20b}).

\begin{acknowledgments}
SEIB is funded by the Deutsche Forschungsgemeinschaft (DFG) under Emmy Noether grant number BO 5771/1-1. GK is partly supported by the Department of Atomic Energy (Government of India) research project with Project Identification Number RTI~4002, and by the Max Planck Society through a Max Planck Partner Group. For the purpose of open access, the authors have applied a Creative Commons Attribution (CC BY) licence to any Author Accepted Manuscript version arising from this submission.
\end{acknowledgments}

\bibliographystyle{aasjournal}

\begin{thebibliography}{}
\expandafter\ifx\csname natexlab\endcsname\relax\def\natexlab#1{#1}\fi
\providecommand{\url}[1]{\href{#1}{#1}}
\providecommand{\dodoi}[1]{doi:~\href{http://doi.org/#1}{\nolinkurl{#1}}}
\providecommand{\doeprint}[1]{\href{http://ascl.net/#1}{\nolinkurl{http://ascl.net/#1}}}
\providecommand{\doarXiv}[1]{\href{https://arxiv.org/abs/#1}{\nolinkurl{https://arxiv.org/abs/#1}}}

\bibitem[{{Abel} \& {Haehnelt}(1999)}]{AH99}
{Abel}, T., \& {Haehnelt}, M.~G. 1999, \apjl, 520, L13, \dodoi{10.1086/312136}

\bibitem[{{Almgren} {et~al.}(2013){Almgren}, {Bell}, {Lijewski}, {Luki{\'c}},
  \& {Van Andel}}]{Almgren13}
{Almgren}, A.~S., {Bell}, J.~B., {Lijewski}, M.~J., {Luki{\'c}}, Z., \& {Van
  Andel}, E. 2013, \apj, 765, 39, \dodoi{10.1088/0004-637X/765/1/39}

\bibitem[{{Alsing} {et~al.}(2018){Alsing}, {Wandelt}, \& {Feeney}}]{Alsing18}
{Alsing}, J., {Wandelt}, B., \& {Feeney}, S. 2018, \mnras, 477, 2874,
  \dodoi{10.1093/mnras/sty819}

\bibitem[{{Becker} \& {Bolton}(2013)}]{BB13}
{Becker}, G.~D., \& {Bolton}, J.~S. 2013, \mnras, 436, 1023,
  \dodoi{10.1093/mnras/stt1610}

\bibitem[{{Becker} {et~al.}(2015){Becker}, {Bolton}, {Madau}, {Pettini},
  {Ryan-Weber}, \& {Venemans}}]{Becker15}
{Becker}, G.~D., {Bolton}, J.~S., {Madau}, P., {et~al.} 2015, \mnras, 447,
  3402, \dodoi{10.1093/mnras/stu2646}

\bibitem[{{Becker} {et~al.}(2021){Becker}, {D'Aloisio}, {Christenson}, {Zhu},
  {Worseck}, \& {Bolton}}]{Becker21}
{Becker}, G.~D., {D'Aloisio}, A., {Christenson}, H.~M., {et~al.} 2021, \mnras,
  508, 1853, \dodoi{10.1093/mnras/stab2696}

\bibitem[{{Becker} {et~al.}(2018){Becker}, {Davies}, {Furlanetto}, {Malkan},
  {Boera}, \& {Douglass}}]{Becker18}
{Becker}, G.~D., {Davies}, F.~B., {Furlanetto}, S.~R., {et~al.} 2018, \apj,
  863, 92, \dodoi{10.3847/1538-4357/aacc73}

\bibitem[{{Becker} {et~al.}(2001){Becker}, {Fan}, {White}, {Strauss},
  {Narayanan}, {Lupton}, {Gunn}, {Annis}, {Bahcall}, {Brinkmann}, {Connolly},
  {Csabai}, {Czarapata}, {Doi}, {Heckman}, {Hennessy}, {Ivezi{\'c}}, {Knapp},
  {Lamb}, {McKay}, {Munn}, {Nash}, {Nichol}, {Pier}, {Richards}, {Schneider},
  {Stoughton}, {Szalay}, {Thakar}, \& {York}}]{Becker01}
{Becker}, R.~H., {Fan}, X., {White}, R.~L., {et~al.} 2001, \aj, 122, 2850,
  \dodoi{10.1086/324231}

\bibitem[{Blum(2010)}]{Blum10}
Blum, M. G.~B. 2010, Journal of the American Statistical Association, 105,
  1178, \dodoi{10.1198/jasa.2010.tm09448}

\bibitem[{{Boera} {et~al.}(2019){Boera}, {Becker}, {Bolton}, \&
  {Nasir}}]{Boera19}
{Boera}, E., {Becker}, G.~D., {Bolton}, J.~S., \& {Nasir}, F. 2019, \apj, 872,
  101, \dodoi{10.3847/1538-4357/aafee4}

\bibitem[{{Bolton} {et~al.}(2012){Bolton}, {Becker}, {Raskutti}, {Wyithe},
  {Haehnelt}, \& {Sargent}}]{Bolton12}
{Bolton}, J.~S., {Becker}, G.~D., {Raskutti}, S., {et~al.} 2012, \mnras, 419,
  2880, \dodoi{10.1111/j.1365-2966.2011.19929.x}

\bibitem[{{Bolton} \& {Haehnelt}(2007)}]{BH07a}
{Bolton}, J.~S., \& {Haehnelt}, M.~G. 2007, \mnras, 382, 325,
  \dodoi{10.1111/j.1365-2966.2007.12372.x}

\bibitem[{{Bolton} {et~al.}(2005){Bolton}, {Haehnelt}, {Viel}, \&
  {Springel}}]{Bolton05}
{Bolton}, J.~S., {Haehnelt}, M.~G., {Viel}, M., \& {Springel}, V. 2005, \mnras,
  357, 1178, \dodoi{10.1111/j.1365-2966.2005.08704.x}

\bibitem[{{Bolton} {et~al.}(2017){Bolton}, {Puchwein}, {Sijacki}, {Haehnelt},
  {Kim}, {Meiksin}, {Regan}, \& {Viel}}]{Bolton17}
{Bolton}, J.~S., {Puchwein}, E., {Sijacki}, D., {et~al.} 2017, \mnras, 464,
  897, \dodoi{10.1093/mnras/stw2397}

\bibitem[{{Bosman}(2021)}]{Bosman21MFP}
{Bosman}, S. E.~I. 2021, arXiv e-prints, arXiv:2108.12446.
\newblock \doarXiv{2108.12446}

\bibitem[{{Bosman} {et~al.}(2018){Bosman}, {Fan}, {Jiang}, {Reed}, {Matsuoka},
  {Becker}, \& {Haehnelt}}]{Bosman18}
{Bosman}, S.~E.~I., {Fan}, X., {Jiang}, L., {et~al.} 2018, \mnras, 479, 1055,
  \dodoi{10.1093/mnras/sty1344}

\bibitem[{{Bosman} {et~al.}(2020){Bosman}, {{\v{D}}urov{\v{c}}{\'\i}kov{\'a}},
  {Davies}, \& {Eilers}}]{Bosman20b}
{Bosman}, S.~E.~I., {{\v{D}}urov{\v{c}}{\'\i}kov{\'a}}, D., {Davies}, F.~B., \&
  {Eilers}, A.~C. 2020, arXiv e-prints, arXiv:2006.10744.
\newblock \doarXiv{2006.10744}

\bibitem[{{Bosman} {et~al.}(2022){Bosman}, {Davies}, {Becker}, {Keating},
  {Davies}, {Zhu}, {Eilers}, {D'Odorico}, {Bian}, {Bischetti}, {Cristiani},
  {Fan}, {Farina}, {Haehnelt}, {Hennawi}, {Kulkarni}, {Mesinger}, {Meyer},
  {Onoue}, {Pallottini}, {Qin}, {Ryan-Weber}, {Schindler}, {Walter}, {Wang}, \&
  {Yang}}]{Bosman22}
{Bosman}, S. E.~I., {Davies}, F.~B., {Becker}, G.~D., {et~al.} 2022, \mnras,
  514, 55, \dodoi{10.1093/mnras/stac1046}

\bibitem[{{Bouwens} {et~al.}(2015{\natexlab{a}}){Bouwens}, {Illingworth},
  {Oesch}, {Caruana}, {Holwerda}, {Smit}, \& {Wilkins}}]{Bouwens15b}
{Bouwens}, R.~J., {Illingworth}, G.~D., {Oesch}, P.~A., {et~al.}
  2015{\natexlab{a}}, \apj, 811, 140, \dodoi{10.1088/0004-637X/811/2/140}

\bibitem[{{Bouwens} {et~al.}(2015{\natexlab{b}}){Bouwens}, {Illingworth},
  {Oesch}, {Trenti}, {Labb{\'e}}, {Bradley}, {Carollo}, {van Dokkum},
  {Gonzalez}, {Holwerda}, {Franx}, {Spitler}, {Smit}, \& {Magee}}]{Bouwens15}
---. 2015{\natexlab{b}}, \apj, 803, 34, \dodoi{10.1088/0004-637X/803/1/34}

\bibitem[{{Cain} {et~al.}(2021){Cain}, {D'Aloisio}, {Gangolli}, \&
  {Becker}}]{Cain21}
{Cain}, C., {D'Aloisio}, A., {Gangolli}, N., \& {Becker}, G.~D. 2021, \apjl,
  917, L37, \dodoi{10.3847/2041-8213/ac1ace}

\bibitem[{{Calverley} {et~al.}(2011){Calverley}, {Becker}, {Haehnelt}, \&
  {Bolton}}]{Calverley11}
{Calverley}, A.~P., {Becker}, G.~D., {Haehnelt}, M.~G., \& {Bolton}, J.~S.
  2011, \mnras, 412, 2543, \dodoi{10.1111/j.1365-2966.2010.18072.x}

\bibitem[{{Chardin} {et~al.}(2015){Chardin}, {Haehnelt}, {Aubert}, \&
  {Puchwein}}]{Chardin15}
{Chardin}, J., {Haehnelt}, M.~G., {Aubert}, D., \& {Puchwein}, E. 2015, \mnras,
  453, 2943, \dodoi{10.1093/mnras/stv1786}

\bibitem[{{Chardin} {et~al.}(2017){Chardin}, {Puchwein}, \&
  {Haehnelt}}]{Chardin17}
{Chardin}, J., {Puchwein}, E., \& {Haehnelt}, M.~G. 2017, \mnras, 465, 3429,
  \dodoi{10.1093/mnras/stw2943}

\bibitem[{{Choudhury} {et~al.}(2021){Choudhury}, {Paranjape}, \&
  {Bosman}}]{Choudhury21}
{Choudhury}, T.~R., {Paranjape}, A., \& {Bosman}, S. E.~I. 2021, \mnras, 501,
  5782, \dodoi{10.1093/mnras/stab045}

\bibitem[{{Christenson} {et~al.}(2021){Christenson}, {Becker}, {Furlanetto},
  {Davies}, {Malkan}, {Zhu}, {Boera}, \& {Trapp}}]{Christenson21}
{Christenson}, H.~M., {Becker}, G.~D., {Furlanetto}, S.~R., {et~al.} 2021,
  \apj, 923, 87, \dodoi{10.3847/1538-4357/ac2a34}

\bibitem[{{Cole} {et~al.}(2022){Cole}, {Miller}, {Witte}, {Cai}, {Grootes},
  {Nattino}, \& {Weniger}}]{Cole22}
{Cole}, A., {Miller}, B.~K., {Witte}, S.~J., {et~al.} 2022, \jcap, 2022, 004,
  \dodoi{10.1088/1475-7516/2022/09/004}

\bibitem[{{D'Aloisio} {et~al.}(2018){D'Aloisio}, {McQuinn}, {Davies}, \&
  {Furlanetto}}]{D'Aloisio18}
{D'Aloisio}, A., {McQuinn}, M., {Davies}, F.~B., \& {Furlanetto}, S.~R. 2018,
  \mnras, 473, 560, \dodoi{10.1093/mnras/stx2341}

\bibitem[{{D'Aloisio} {et~al.}(2019){D'Aloisio}, {McQuinn}, {Maupin}, {Davies},
  {Trac}, {Fuller}, \& {Upton Sanderbeck}}]{D'Aloisio18b}
{D'Aloisio}, A., {McQuinn}, M., {Maupin}, O., {et~al.} 2019, \apj, 874, 154,
  \dodoi{10.3847/1538-4357/ab0d83}

\bibitem[{{D'Aloisio} {et~al.}(2015){D'Aloisio}, {McQuinn}, \&
  {Trac}}]{D'Aloisio15}
{D'Aloisio}, A., {McQuinn}, M., \& {Trac}, H. 2015, \apjl, 813, L38,
  \dodoi{10.1088/2041-8205/813/2/L38}

\bibitem[{{D'Aloisio} {et~al.}(2020){D'Aloisio}, {McQuinn}, {Trac}, {Cain}, \&
  {Mesinger}}]{D'Aloisio20}
{D'Aloisio}, A., {McQuinn}, M., {Trac}, H., {Cain}, C., \& {Mesinger}, A. 2020,
  \apj, 898, 149, \dodoi{10.3847/1538-4357/ab9f2f}

\bibitem[{{D'Aloisio} {et~al.}(2017){D'Aloisio}, {Upton Sanderbeck}, {McQuinn},
  {Trac}, \& {Shapiro}}]{D'Aloisio17}
{D'Aloisio}, A., {Upton Sanderbeck}, P.~R., {McQuinn}, M., {Trac}, H., \&
  {Shapiro}, P.~R. 2017, \mnras, 468, 4691, \dodoi{10.1093/mnras/stx711}

\bibitem[{{Davies} {et~al.}(2018{\natexlab{a}}){Davies}, {Becker}, \&
  {Furlanetto}}]{Davies17b}
{Davies}, F.~B., {Becker}, G.~D., \& {Furlanetto}, S.~R. 2018{\natexlab{a}},
  \apj, 860, 155, \dodoi{10.3847/1538-4357/aac2d6}

\bibitem[{{Davies} {et~al.}(2021){Davies}, {Bosman}, {Furlanetto}, {Becker}, \&
  {D'Aloisio}}]{Davies21}
{Davies}, F.~B., {Bosman}, S. E.~I., {Furlanetto}, S.~R., {Becker}, G.~D., \&
  {D'Aloisio}, A. 2021, \apjl, 918, L35, \dodoi{10.3847/2041-8213/ac1ffb}

\bibitem[{{Davies} \& {Furlanetto}(2016)}]{DF16}
{Davies}, F.~B., \& {Furlanetto}, S.~R. 2016, \mnras, 460, 1328,
  \dodoi{10.1093/mnras/stw931}

\bibitem[{{Davies} {et~al.}(2018{\natexlab{b}}){Davies}, {Hennawi}, {Eilers},
  \& {Luki{\'c}}}]{Davies17}
{Davies}, F.~B., {Hennawi}, J.~F., {Eilers}, A.-C., \& {Luki{\'c}}, Z.
  2018{\natexlab{b}}, \apj, 855, 106, \dodoi{10.3847/1538-4357/aaaf70}

\bibitem[{{Davies} {et~al.}(2018{\natexlab{c}}){Davies}, {Hennawi},
  {Ba{\~n}ados}, {Luki{\'c}}, {Decarli}, {Fan}, {Farina}, {Mazzucchelli},
  {Rix}, {Venemans}, {Walter}, {Wang}, \& {Yang}}]{Davies18b}
{Davies}, F.~B., {Hennawi}, J.~F., {Ba{\~n}ados}, E., {et~al.}
  2018{\natexlab{c}}, \apj, 864, 142, \dodoi{10.3847/1538-4357/aad6dc}

\bibitem[{{Davies} {et~al.}(2018{\natexlab{d}}){Davies}, {Hennawi},
  {Ba{\~n}ados}, {Simcoe}, {Decarli}, {Fan}, {Farina}, {Mazzucchelli}, {Rix},
  {Venemans}, {Walter}, {Wang}, \& {Yang}}]{Davies18a}
---. 2018{\natexlab{d}}, \apj, 864, 143, \dodoi{10.3847/1538-4357/aad7f8}

\bibitem[{{Davies} {et~al.}(2023){Davies}, {Ryan-Weber}, {D'Odorico}, {Bosman},
  {Meyer}, {Becker}, {Cupani}, {Keating}, {Bischetti}, {Davies}, {Eilers},
  {Farina}, {Haehnelt}, {Pallottini}, \& {Zhu}}]{RDavies23}
{Davies}, R.~L., {Ryan-Weber}, E., {D'Odorico}, V., {et~al.} 2023, \mnras, 521,
  314, \dodoi{10.1093/mnras/stad294}

\bibitem[{{D'Odorico} {et~al.}(2023){D'Odorico}, {Ba{\~n}ados}, {Becker},
  {Bischetti}, {Bosman}, {Cupani}, {Davies}, {Farina}, {Ferrara}, {Feruglio},
  {Mazzucchelli}, {Ryan-Weber}, {Schindler}, {Sodini}, {Venemans}, {Walter},
  {Chen}, {Lai}, {Zhu}, {Bian}, {Campo}, {Carniani}, {Cristiani}, {Davies},
  {Decarli}, {Drake}, {Eilers}, {Fan}, {Gaikwad}, {Gallerani}, {Greig},
  {Haehnelt}, {Hennawi}, {Keating}, {Kulkarni}, {Mesinger}, {Meyer},
  {Neeleman}, {Onoue}, {Pallottini}, {Qin}, {Rojas-Ruiz}, {Satyavolu},
  {Sebastian}, {Tripodi}, {Wang}, {Wolfson}, {Yang}, \&
  {Zanchettin}}]{D'Odorico23}
{D'Odorico}, V., {Ba{\~n}ados}, E., {Becker}, G.~D., {et~al.} 2023, \mnras,
  523, 1399, \dodoi{10.1093/mnras/stad1468}

\bibitem[{{Eilers} {et~al.}(2018){Eilers}, {Davies}, \& {Hennawi}}]{Eilers18}
{Eilers}, A.-C., {Davies}, F.~B., \& {Hennawi}, J.~F. 2018, \apj, 864, 53,
  \dodoi{10.3847/1538-4357/aad4fd}

\bibitem[{{Emberson} {et~al.}(2013){Emberson}, {Thomas}, \&
  {Alvarez}}]{Emberson13}
{Emberson}, J.~D., {Thomas}, R.~M., \& {Alvarez}, M.~A. 2013, \apj, 763, 146,
  \dodoi{10.1088/0004-637X/763/2/146}

\bibitem[{{Fan} {et~al.}(2006){Fan}, {Strauss}, {Becker}, {White}, {Gunn},
  {Knapp}, {Richards}, {Schneider}, {Brinkmann}, \& {Fukugita}}]{Fan06}
{Fan}, X., {Strauss}, M.~A., {Becker}, R.~H., {et~al.} 2006, \aj, 132, 117,
  \dodoi{10.1086/504836}

\bibitem[{{Faucher-Gigu{\`e}re}(2020)}]{FG20}
{Faucher-Gigu{\`e}re}, C.-A. 2020, \mnras, 493, 1614,
  \dodoi{10.1093/mnras/staa302}

\bibitem[{{Fumagalli} {et~al.}(2013){Fumagalli}, {O'Meara}, {Prochaska}, \&
  {Worseck}}]{Fumagalli13}
{Fumagalli}, M., {O'Meara}, J.~M., {Prochaska}, J.~X., \& {Worseck}, G. 2013,
  \apj, 775, 78, \dodoi{10.1088/0004-637X/775/1/78}

\bibitem[{{Furlanetto} {et~al.}(2004){Furlanetto}, {Zaldarriaga}, \&
  {Hernquist}}]{Furlanetto04}
{Furlanetto}, S.~R., {Zaldarriaga}, M., \& {Hernquist}, L. 2004, \apj, 613, 1,
  \dodoi{10.1086/423025}

\bibitem[{{Gaikwad} {et~al.}(2021){Gaikwad}, {Srianand}, {Haehnelt}, \&
  {Choudhury}}]{Gaikwad21}
{Gaikwad}, P., {Srianand}, R., {Haehnelt}, M.~G., \& {Choudhury}, T.~R. 2021,
  \mnras, \dodoi{10.1093/mnras/stab2017}

\bibitem[{{Gaikwad} {et~al.}(2020){Gaikwad}, {Rauch}, {Haehnelt}, {Puchwein},
  {Bolton}, {Keating}, {Kulkarni}, {Ir{\v{s}}i{\v{c}}}, {Ba{\~n}ados},
  {Becker}, {Boera}, {Zahedy}, {Chen}, {Carswell}, {Chardin}, \&
  {Rorai}}]{Gaikwad20}
{Gaikwad}, P., {Rauch}, M., {Haehnelt}, M.~G., {et~al.} 2020, \mnras,
  \dodoi{10.1093/mnras/staa907}

\bibitem[{{Gaikwad} {et~al.}(2023){Gaikwad}, {Haehnelt}, {Davies}, {Bosman},
  {Molaro}, {Kulkarni}, {D'Odorico}, {Becker}, {Davies}, {Nasir}, {Bolton},
  {Keating}, {Ir{\v{s}}i{\v{c}}}, {Puchwein}, {Zhu}, {Asthana}, {Yang}, {Lai},
  \& {Eilers}}]{Gaikwad23}
{Gaikwad}, P., {Haehnelt}, M.~G., {Davies}, F.~B., {et~al.} 2023, arXiv
  e-prints, arXiv:2304.02038, \dodoi{10.48550/arXiv.2304.02038}

\bibitem[{{Garaldi} {et~al.}(2022){Garaldi}, {Kannan}, {Smith}, {Springel},
  {Pakmor}, {Vogelsberger}, \& {Hernquist}}]{Garaldi22}
{Garaldi}, E., {Kannan}, R., {Smith}, A., {et~al.} 2022, \mnras, 512, 4909,
  \dodoi{10.1093/mnras/stac257}

\bibitem[{{Giallongo} {et~al.}(2015){Giallongo}, {Grazian}, {Fiore}, {Fontana},
  {Pentericci}, {Vanzella}, {Dickinson}, {Kocevski}, {Castellano}, {Cristiani},
  {Ferguson}, {Finkelstein}, {Grogin}, {Hathi}, {Koekemoer}, {Newman}, \&
  {Salvato}}]{Giallongo15}
{Giallongo}, E., {Grazian}, A., {Fiore}, F., {et~al.} 2015, \aap, 578, A83,
  \dodoi{10.1051/0004-6361/201425334}

\bibitem[{{Gnedin}(2000)}]{Gnedin00}
{Gnedin}, N.~Y. 2000, \apj, 542, 535, \dodoi{10.1086/317042}

\bibitem[{{Gnedin} \& {Hui}(1998)}]{GH98}
{Gnedin}, N.~Y., \& {Hui}, L. 1998, \mnras, 296, 44,
  \dodoi{10.1046/j.1365-8711.1998.01249.x}

\bibitem[{{Greig} {et~al.}(2022){Greig}, {Mesinger}, {Davies}, {Wang}, {Yang},
  \& {Hennawi}}]{Greig22}
{Greig}, B., {Mesinger}, A., {Davies}, F.~B., {et~al.} 2022, \mnras, 512, 5390,
  \dodoi{10.1093/mnras/stac825}

\bibitem[{{Gunn} \& {Peterson}(1965)}]{GP65}
{Gunn}, J.~E., \& {Peterson}, B.~A. 1965, \apj, 142, 1633,
  \dodoi{10.1086/148444}

\bibitem[{{Haardt} \& {Madau}(2012)}]{HM12}
{Haardt}, F., \& {Madau}, P. 2012, \apj, 746, 125,
  \dodoi{10.1088/0004-637X/746/2/125}

\bibitem[{{Harikane} {et~al.}(2023){Harikane}, {Zhang}, {Nakajima}, {Ouchi},
  {Isobe}, {Ono}, {Hatano}, {Xu}, \& {Umeda}}]{Harikane23}
{Harikane}, Y., {Zhang}, Y., {Nakajima}, K., {et~al.} 2023, arXiv e-prints,
  arXiv:2303.11946, \dodoi{10.48550/arXiv.2303.11946}

\bibitem[{{Hoag} {et~al.}(2019){Hoag}, {Brada{\v{c}}}, {Huang}, {Mason},
  {Treu}, {Schmidt}, {Trenti}, {Strait}, {Lemaux}, {Finney}, \&
  {Paddock}}]{Hoag19}
{Hoag}, A., {Brada{\v{c}}}, M., {Huang}, K., {et~al.} 2019, \apj, 878, 12,
  \dodoi{10.3847/1538-4357/ab1de7}

\bibitem[{{Jung} {et~al.}(2020){Jung}, {Finkelstein}, {Dickinson}, {Hutchison},
  {Larson}, {Papovich}, {Pentericci}, {Straughn}, {Guo}, {Malhotra}, {Rhoads},
  {Song}, {Tilvi}, \& {Wold}}]{Jung20}
{Jung}, I., {Finkelstein}, S.~L., {Dickinson}, M., {et~al.} 2020, \apj, 904,
  144, \dodoi{10.3847/1538-4357/abbd44}

\bibitem[{{Kashino} {et~al.}(2020){Kashino}, {Lilly}, {Shibuya}, {Ouchi}, \&
  {Kashikawa}}]{Kashino20}
{Kashino}, D., {Lilly}, S.~J., {Shibuya}, T., {Ouchi}, M., \& {Kashikawa}, N.
  2020, \apj, 888, 6, \dodoi{10.3847/1538-4357/ab5a7d}

\bibitem[{{Keating} {et~al.}(2018){Keating}, {Puchwein}, \&
  {Haehnelt}}]{Keating18}
{Keating}, L.~C., {Puchwein}, E., \& {Haehnelt}, M.~G. 2018, \mnras, 477, 5501,
  \dodoi{10.1093/mnras/sty968}

\bibitem[{{Keating} {et~al.}(2020){Keating}, {Weinberger}, {Kulkarni},
  {Haehnelt}, {Chardin}, \& {Aubert}}]{Keating19}
{Keating}, L.~C., {Weinberger}, L.~H., {Kulkarni}, G., {et~al.} 2020, \mnras,
  491, 1736, \dodoi{10.1093/mnras/stz3083}

\bibitem[{{Khaire} \& {Srianand}(2019)}]{KS19}
{Khaire}, V., \& {Srianand}, R. 2019, \mnras, 484, 4174,
  \dodoi{10.1093/mnras/stz174}

\bibitem[{{Kocevski} {et~al.}(2023){Kocevski}, {Onoue}, {Inayoshi}, {Trump},
  {Arrabal Haro}, {Grazian}, {Dickinson}, {Finkelstein}, {Kartaltepe},
  {Hirschmann}, {Fujimoto}, {Juneau}, {Amorin}, {Bagley}, {Barro}, {Bell},
  {Bisigello}, {Calabro}, {Cleri}, {Cooper}, {Ding}, {Grogin}, {Ho}, {Inoue},
  {Jiang}, {Jones}, {Koekemoer}, {Li}, {Li}, {McGrath}, {Molina}, {Papovich},
  {Perez-Gonzalez}, {Pirzkal}, {Wilkins}, {Yang}, \& {Yung}}]{Kocevski23}
{Kocevski}, D.~D., {Onoue}, M., {Inayoshi}, K., {et~al.} 2023, arXiv e-prints,
  arXiv:2302.00012, \dodoi{10.48550/arXiv.2302.00012}

\bibitem[{{Kulkarni} {et~al.}(2015){Kulkarni}, {Hennawi}, {O{\~n}orbe},
  {Rorai}, \& {Springel}}]{Kulkarni15}
{Kulkarni}, G., {Hennawi}, J.~F., {O{\~n}orbe}, J., {Rorai}, A., \& {Springel},
  V. 2015, \apj, 812, 30, \dodoi{10.1088/0004-637X/812/1/30}

\bibitem[{{Kulkarni} {et~al.}(2019){Kulkarni}, {Keating}, {Haehnelt}, {Bosman},
  {Puchwein}, {Chardin}, \& {Aubert}}]{Kulkarni19}
{Kulkarni}, G., {Keating}, L.~C., {Haehnelt}, M.~G., {et~al.} 2019, \mnras,
  485, L24, \dodoi{10.1093/mnrasl/slz025}

\bibitem[{{Labbe} {et~al.}(2023){Labbe}, {Greene}, {Bezanson}, {Fujimoto},
  {Furtak}, {Goulding}, {Matthee}, {Naidu}, {Oesch}, {Atek}, {Brammer},
  {Chemerynska}, {Coe}, {Cutler}, {Dayal}, {Feldmann}, {Franx}, {Glazebrook},
  {Leja}, {Marchesini}, {Maseda}, {Nanayakkara}, {Nelson}, {Pan}, {Papovich},
  {Price}, {Suess}, {Wang}, {Whitaker}, {Williams}, \& {Zitrin}}]{Labbe23}
{Labbe}, I., {Greene}, J.~E., {Bezanson}, R., {et~al.} 2023, arXiv e-prints,
  arXiv:2306.07320, \dodoi{10.48550/arXiv.2306.07320}

\bibitem[{{Lewis} {et~al.}(2022){Lewis}, {Ocvirk}, {Sorce}, {Dubois}, {Aubert},
  {Conaboy}, {Shapiro}, {Dawoodbhoy}, {Teyssier}, {Yepes}, {Gottl{\"o}ber},
  {Rasera}, {Ahn}, {Iliev}, {Park}, \& {Th{\'e}lie}}]{Lewis22}
{Lewis}, J. S.~W., {Ocvirk}, P., {Sorce}, J.~G., {et~al.} 2022, \mnras, 516,
  3389, \dodoi{10.1093/mnras/stac2383}

\bibitem[{{Lidz} {et~al.}(2006){Lidz}, {Oh}, \& {Furlanetto}}]{Lidz06b}
{Lidz}, A., {Oh}, S.~P., \& {Furlanetto}, S.~R. 2006, \apjl, 639, L47,
  \dodoi{10.1086/502678}

\bibitem[{{Luki{\'c}} {et~al.}(2007){Luki{\'c}}, {Heitmann}, {Habib},
  {Bashinsky}, \& {Ricker}}]{Lukic07}
{Luki{\'c}}, Z., {Heitmann}, K., {Habib}, S., {Bashinsky}, S., \& {Ricker},
  P.~M. 2007, \apj, 671, 1160, \dodoi{10.1086/523083}

\bibitem[{{Luki{\'c}} {et~al.}(2015){Luki{\'c}}, {Stark}, {Nugent}, {White},
  {Meiksin}, \& {Almgren}}]{Lukic15}
{Luki{\'c}}, Z., {Stark}, C.~W., {Nugent}, P., {et~al.} 2015, \mnras, 446,
  3697, \dodoi{10.1093/mnras/stu2377}

\bibitem[{{Maiolino} {et~al.}(2023){Maiolino}, {Scholtz}, {Curtis-Lake},
  {Carniani}, {Baker}, {de Graaff}, {Tacchella}, {{\"U}bler}, {D'Eugenio},
  {Witstok}, {Curti}, {Arribas}, {Bunker}, {Charlot}, {Chevallard},
  {Eisenstein}, {Egami}, {Ji}, {Jones}, {Lyu}, {Rawle}, {Robertson},
  {Rujopakarn}, {Perna}, {Sun}, {Venturi}, {Williams}, \&
  {Willott}}]{Maiolino23}
{Maiolino}, R., {Scholtz}, J., {Curtis-Lake}, E., {et~al.} 2023, arXiv
  e-prints, arXiv:2308.01230, \dodoi{10.48550/arXiv.2308.01230}

\bibitem[{Marin {et~al.}(2012)Marin, Pudlo, Robert, \& Ryder}]{Marin12}
Marin, J.-M., Pudlo, P., Robert, C.~P., \& Ryder, R.~J. 2012, Statistics and
  Computing, 22, 1167, \dodoi{10.1007/s11222-011-9288-2}

\bibitem[{{Mason} {et~al.}(2018){Mason}, {Treu}, {Dijkstra}, {Mesinger},
  {Trenti}, {Pentericci}, {de Barros}, \& {Vanzella}}]{Mason18}
{Mason}, C.~A., {Treu}, T., {Dijkstra}, M., {et~al.} 2018, \apj, 856, 2,
  \dodoi{10.3847/1538-4357/aab0a7}

\bibitem[{{Mason} {et~al.}(2019){Mason}, {Fontana}, {Treu}, {Schmidt}, {Hoag},
  {Abramson}, {Amorin}, {Brada{\v{c}}}, {Guaita}, {Jones}, {Henry}, {Malkan},
  {Pentericci}, {Trenti}, \& {Vanzella}}]{Mason19}
{Mason}, C.~A., {Fontana}, A., {Treu}, T., {et~al.} 2019, \mnras, 485, 3947,
  \dodoi{10.1093/mnras/stz632}

\bibitem[{{Matsuoka} {et~al.}(2018){Matsuoka}, {Strauss}, {Kashikawa}, {Onoue},
  {Iwasawa}, {Tang}, {Lee}, {Imanishi}, {Nagao}, {Akiyama}, {Asami}, {Bosch},
  {Furusawa}, {Goto}, {Gunn}, {Harikane}, {Ikeda}, {Izumi}, {Kawaguchi},
  {Kato}, {Kikuta}, {Kohno}, {Komiyama}, {Lupton}, {Minezaki}, {Miyazaki},
  {Murayama}, {Niida}, {Nishizawa}, {Noboriguchi}, {Oguri}, {Ono}, {Ouchi},
  {Price}, {Sameshima}, {Schulze}, {Shirakata}, {Silverman}, {Sugiyama},
  {Tait}, {Takada}, {Takata}, {Tanaka}, {Toba}, {Utsumi}, {Wang}, \&
  {Yamashita}}]{Matsuoka18}
{Matsuoka}, Y., {Strauss}, M.~A., {Kashikawa}, N., {et~al.} 2018, \apj, 869,
  150, \dodoi{10.3847/1538-4357/aaee7a}

\bibitem[{{Matthee} {et~al.}(2023){Matthee}, {Naidu}, {Brammer}, {Chisholm},
  {Eilers}, {Goulding}, {Greene}, {Kashino}, {Labbe}, {Lilly}, {Mackenzie},
  {Oesch}, {Weibel}, {Wuyts}, {Xiao}, {Bordoloi}, {Bouwens}, {van Dokkum},
  {Illingworth}, {Kramarenko}, {Maseda}, {Mason}, {Meyer}, {Nelson}, {Reddy},
  {Shivaei}, {Simcoe}, \& {Yue}}]{Matthee23}
{Matthee}, J., {Naidu}, R.~P., {Brammer}, G., {et~al.} 2023, arXiv e-prints,
  arXiv:2306.05448, \dodoi{10.48550/arXiv.2306.05448}

\bibitem[{{McQuinn} {et~al.}(2011){McQuinn}, {Oh}, \&
  {Faucher-Gigu{\`e}re}}]{McQuinn11}
{McQuinn}, M., {Oh}, S.~P., \& {Faucher-Gigu{\`e}re}, C.-A. 2011, \apj, 743,
  82, \dodoi{10.1088/0004-637X/743/1/82}

\bibitem[{{Meiksin}(2020)}]{Meiksin20}
{Meiksin}, A. 2020, \mnras, 491, 4884, \dodoi{10.1093/mnras/stz3395}

\bibitem[{{Meiksin} \& {White}(2003)}]{MW03}
{Meiksin}, A., \& {White}, M. 2003, \mnras, 342, 1205,
  \dodoi{10.1046/j.1365-8711.2003.06624.x}

\bibitem[{{Mesinger} \& {Furlanetto}(2007)}]{MF07}
{Mesinger}, A., \& {Furlanetto}, S. 2007, \apj, 669, 663,
  \dodoi{10.1086/521806}

\bibitem[{{Mesinger} \& {Furlanetto}(2009)}]{MF09}
---. 2009, \mnras, 400, 1461, \dodoi{10.1111/j.1365-2966.2009.15547.x}

\bibitem[{{Mesinger} {et~al.}(2011){Mesinger}, {Furlanetto}, \&
  {Cen}}]{Mesinger11}
{Mesinger}, A., {Furlanetto}, S., \& {Cen}, R. 2011, \mnras, 411, 955,
  \dodoi{10.1111/j.1365-2966.2010.17731.x}

\bibitem[{{Miralda-Escud{\'e}} \& {Rees}(1994)}]{MR94}
{Miralda-Escud{\'e}}, J., \& {Rees}, M.~J. 1994, \mnras, 266, 343,
  \dodoi{10.1093/mnras/266.2.343}

\bibitem[{{Mu{\~n}oz} {et~al.}(2016){Mu{\~n}oz}, {Oh}, {Davies}, \&
  {Furlanetto}}]{Munoz16}
{Mu{\~n}oz}, J.~A., {Oh}, S.~P., {Davies}, F.~B., \& {Furlanetto}, S.~R. 2016,
  \mnras, 455, 1385, \dodoi{10.1093/mnras/stv2355}

\bibitem[{{Nasir} {et~al.}(2016){Nasir}, {Bolton}, \& {Becker}}]{Nasir16}
{Nasir}, F., {Bolton}, J.~S., \& {Becker}, G.~D. 2016, \mnras, 463, 2335,
  \dodoi{10.1093/mnras/stw2147}

\bibitem[{{Nasir} \& {D'Aloisio}(2020)}]{ND20}
{Nasir}, F., \& {D'Aloisio}, A. 2020, \mnras, 494, 3080,
  \dodoi{10.1093/mnras/staa894}

\bibitem[{{O{\~n}orbe} {et~al.}(2019){O{\~n}orbe}, {Davies}, {Luki{\'c}}, {},
  {Hennawi}, \& {Sorini}}]{Onorbe18}
{O{\~n}orbe}, J., {Davies}, F.~B., {Luki{\'c}}, {et~al.} 2019, \mnras, 486,
  4075, \dodoi{10.1093/mnras/stz984}

\bibitem[{{O{\~n}orbe} {et~al.}(2017){O{\~n}orbe}, {Hennawi}, \&
  {Luki{\'c}}}]{Onorbe17}
{O{\~n}orbe}, J., {Hennawi}, J.~F., \& {Luki{\'c}}, Z. 2017, \apj, 837, 106,
  \dodoi{10.3847/1538-4357/aa6031}

\bibitem[{{O'Leary} \& {McQuinn}(2012)}]{OM12}
{O'Leary}, R.~M., \& {McQuinn}, M. 2012, \apj, 760, 4,
  \dodoi{10.1088/0004-637X/760/1/4}

\bibitem[{{Park} {et~al.}(2016){Park}, {Shapiro}, {Choi}, {Yoshida}, {Hirano},
  \& {Ahn}}]{Park16}
{Park}, H., {Shapiro}, P.~R., {Choi}, J.-h., {et~al.} 2016, \apj, 831, 86,
  \dodoi{10.3847/0004-637X/831/1/86}

\bibitem[{{Parsa} {et~al.}(2018){Parsa}, {Dunlop}, \& {McLure}}]{Parsa18}
{Parsa}, S., {Dunlop}, J.~S., \& {McLure}, R.~J. 2018, \mnras, 474, 2904,
  \dodoi{10.1093/mnras/stx2887}

\bibitem[{{Peeples} {et~al.}(2010{\natexlab{a}}){Peeples}, {Weinberg},
  {Dav{\'e}}, {Fardal}, \& {Katz}}]{Peeples10a}
{Peeples}, M.~S., {Weinberg}, D.~H., {Dav{\'e}}, R., {Fardal}, M.~A., \&
  {Katz}, N. 2010{\natexlab{a}}, \mnras, 404, 1281,
  \dodoi{10.1111/j.1365-2966.2010.16383.x}

\bibitem[{{Peeples} {et~al.}(2010{\natexlab{b}}){Peeples}, {Weinberg},
  {Dav{\'e}}, {Fardal}, \& {Katz}}]{Peeples10b}
---. 2010{\natexlab{b}}, \mnras, 404, 1295,
  \dodoi{10.1111/j.1365-2966.2010.16384.x}

\bibitem[{{Planck Collaboration} {et~al.}(2020){Planck Collaboration},
  {Aghanim}, {Akrami}, {Ashdown}, {Aumont}, {Baccigalupi}, {Ballardini},
  {Banday}, {Barreiro}, {Bartolo}, {Basak}, {Battye}, {Benabed}, {Bernard},
  {Bersanelli}, {Bielewicz}, {Bock}, {Bond}, {Borrill}, {Bouchet}, {Boulanger},
  {Bucher}, {Burigana}, {Butler}, {Calabrese}, {Cardoso}, {Carron},
  {Challinor}, {Chiang}, {Chluba}, {Colombo}, {Combet}, {Contreras}, {Crill},
  {Cuttaia}, {de Bernardis}, {de Zotti}, {Delabrouille}, {Delouis}, {Di
  Valentino}, {Diego}, {Dor{\'e}}, {Douspis}, {Ducout}, {Dupac}, {Dusini},
  {Efstathiou}, {Elsner}, {En{\ss}lin}, {Eriksen}, {Fantaye}, {Farhang},
  {Fergusson}, {Fernandez-Cobos}, {Finelli}, {Forastieri}, {Frailis},
  {Fraisse}, {Franceschi}, {Frolov}, {Galeotta}, {Galli}, {Ganga},
  {G{\'e}nova-Santos}, {Gerbino}, {Ghosh}, {Gonz{\'a}lez-Nuevo}, {G{\'o}rski},
  {Gratton}, {Gruppuso}, {Gudmundsson}, {Hamann}, {Handley}, {Hansen},
  {Herranz}, {Hildebrandt}, {Hivon}, {Huang}, {Jaffe}, {Jones}, {Karakci},
  {Keih{\"a}nen}, {Keskitalo}, {Kiiveri}, {Kim}, {Kisner}, {Knox},
  {Krachmalnicoff}, {Kunz}, {Kurki-Suonio}, {Lagache}, {Lamarre}, {Lasenby},
  {Lattanzi}, {Lawrence}, {Le Jeune}, {Lemos}, {Lesgourgues}, {Levrier},
  {Lewis}, {Liguori}, {Lilje}, {Lilley}, {Lindholm}, {L{\'o}pez-Caniego},
  {Lubin}, {Ma}, {Mac{\'\i}as-P{\'e}rez}, {Maggio}, {Maino}, {Mandolesi},
  {Mangilli}, {Marcos-Caballero}, {Maris}, {Martin}, {Martinelli},
  {Mart{\'\i}nez-Gonz{\'a}lez}, {Matarrese}, {Mauri}, {McEwen}, {Meinhold},
  {Melchiorri}, {Mennella}, {Migliaccio}, {Millea}, {Mitra},
  {Miville-Desch{\^e}nes}, {Molinari}, {Montier}, {Morgante}, {Moss}, {Natoli},
  {N{\o}rgaard-Nielsen}, {Pagano}, {Paoletti}, {Partridge}, {Patanchon},
  {Peiris}, {Perrotta}, {Pettorino}, {Piacentini}, {Polastri}, {Polenta},
  {Puget}, {Rachen}, {Reinecke}, {Remazeilles}, {Renzi}, {Rocha}, {Rosset},
  {Roudier}, {Rubi{\~n}o-Mart{\'\i}n}, {Ruiz-Granados}, {Salvati}, {Sandri},
  {Savelainen}, {Scott}, {Shellard}, {Sirignano}, {Sirri}, {Spencer},
  {Sunyaev}, {Suur-Uski}, {Tauber}, {Tavagnacco}, {Tenti}, {Toffolatti},
  {Tomasi}, {Trombetti}, {Valenziano}, {Valiviita}, {Van Tent}, {Vibert},
  {Vielva}, {Villa}, {Vittorio}, {Wandelt}, {Wehus}, {White}, {White},
  {Zacchei}, \& {Zonca}}]{Planck18}
{Planck Collaboration}, {Aghanim}, N., {Akrami}, Y., {et~al.} 2020, \aap, 641,
  A6, \dodoi{10.1051/0004-6361/201833910}

\bibitem[{Pritchard {et~al.}(1999)Pritchard, Seielstad, Perez-Lezaun, \&
  Feldman}]{Pritchard99}
Pritchard, J.~K., Seielstad, M.~T., Perez-Lezaun, A., \& Feldman, M.~W. 1999,
  Molecular Biology and Evolution, 16, 1791

\bibitem[{{Prochaska} {et~al.}(2014){Prochaska}, {Madau}, {O'Meara}, \&
  {Fumagalli}}]{Prochaska14}
{Prochaska}, J.~X., {Madau}, P., {O'Meara}, J.~M., \& {Fumagalli}, M. 2014,
  \mnras, 438, 476, \dodoi{10.1093/mnras/stt2218}

\bibitem[{{Prochaska} {et~al.}(2009){Prochaska}, {Worseck}, \&
  {O'Meara}}]{Prochaska09}
{Prochaska}, J.~X., {Worseck}, G., \& {O'Meara}, J.~M. 2009, \apjl, 705, L113,
  \dodoi{10.1088/0004-637X/705/2/L113}

\bibitem[{{Puchwein} {et~al.}(2019){Puchwein}, {Haardt}, {Haehnelt}, \&
  {Madau}}]{Puchwein19}
{Puchwein}, E., {Haardt}, F., {Haehnelt}, M.~G., \& {Madau}, P. 2019, \mnras,
  485, 47, \dodoi{10.1093/mnras/stz222}

\bibitem[{{Qin} {et~al.}(2021){Qin}, {Mesinger}, {Bosman}, \& {Viel}}]{Qin21}
{Qin}, Y., {Mesinger}, A., {Bosman}, S. E.~I., \& {Viel}, M. 2021, \mnras, 506,
  2390, \dodoi{10.1093/mnras/stab1833}

\bibitem[{{Rahmati} {et~al.}(2013){Rahmati}, {Pawlik}, {Rai{\v c}evi{\'c}}, \&
  {Schaye}}]{Rahmati13}
{Rahmati}, A., {Pawlik}, A.~H., {Rai{\v c}evi{\'c}}, M., \& {Schaye}, J. 2013,
  \mnras, 430, 2427, \dodoi{10.1093/mnras/stt066}

\bibitem[{Rubin(1984)}]{Rubin84}
Rubin, D.~B. 1984, Ann. Statist., 12, 1151, \dodoi{10.1214/aos/1176346785}

\bibitem[{{Rudie} {et~al.}(2013){Rudie}, {Steidel}, {Shapley}, \&
  {Pettini}}]{Rudie13}
{Rudie}, G.~C., {Steidel}, C.~C., {Shapley}, A.~E., \& {Pettini}, M. 2013,
  \apj, 769, 146, \dodoi{10.1088/0004-637X/769/2/146}

\bibitem[{{Sheinis} {et~al.}(2002){Sheinis}, {Bolte}, {Epps}, {Kibrick},
  {Miller}, {Radovan}, {Bigelow}, \& {Sutin}}]{Sheinis02}
{Sheinis}, A.~I., {Bolte}, M., {Epps}, H.~W., {et~al.} 2002, \pasp, 114, 851,
  \dodoi{10.1086/341706}

\bibitem[{Tavar{\'e} {et~al.}(1997)Tavar{\'e}, Balding, Griffiths, \&
  Donnelly}]{Tavare97}
Tavar{\'e}, S., Balding, D.~J., Griffiths, R.~C., \& Donnelly, P. 1997,
  Genetics, 145, 505

\bibitem[{{Tepper-Garc{\'\i}a}(2006)}]{Tepper-Garcia06}
{Tepper-Garc{\'\i}a}, T. 2006, \mnras, 369, 2025,
  \dodoi{10.1111/j.1365-2966.2006.10450.x}

\bibitem[{{Trac} {et~al.}(2008){Trac}, {Cen}, \& {Loeb}}]{Trac08}
{Trac}, H., {Cen}, R., \& {Loeb}, A. 2008, \apjl, 689, L81,
  \dodoi{10.1086/595678}

\bibitem[{{Upton Sanderbeck} {et~al.}(2016){Upton Sanderbeck}, {D'Aloisio}, \&
  {McQuinn}}]{UptonSanderbeck16}
{Upton Sanderbeck}, P.~R., {D'Aloisio}, A., \& {McQuinn}, M.~J. 2016, \mnras,
  460, 1885, \dodoi{10.1093/mnras/stw1117}

\bibitem[{Verner {et~al.}(1996)Verner, Ferland, Korista, \&
  Yakovlev}]{Verner96}
Verner, D.~A., Ferland, G.~J., Korista, K.~T., \& Yakovlev, D.~G. 1996, ApJ,
  465, 487

\bibitem[{{Vernet} {et~al.}(2011){Vernet}, {Dekker}, {D'Odorico}, {Kaper},
  {Kjaergaard}, {Hammer}, {Randich}, {Zerbi}, {Groot}, {Hjorth}, {Guinouard},
  {Navarro}, {Adolfse}, {Albers}, {Amans}, {Andersen}, {Andersen}, {Binetruy},
  {Bristow}, {Castillo}, {Chemla}, {Christensen}, {Conconi}, {Conzelmann},
  {Dam}, {de Caprio}, {de Ugarte Postigo}, {Delabre}, {di Marcantonio},
  {Downing}, {Elswijk}, {Finger}, {Fischer}, {Flores}, {Fran{\c{c}}ois},
  {Goldoni}, {Guglielmi}, {Haigron}, {Hanenburg}, {Hendriks}, {Horrobin},
  {Horville}, {Jessen}, {Kerber}, {Kern}, {Kiekebusch}, {Kleszcz}, {Klougart},
  {Kragt}, {Larsen}, {Lizon}, {Lucuix}, {Mainieri}, {Manuputy}, {Martayan},
  {Mason}, {Mazzoleni}, {Michaelsen}, {Modigliani}, {Moehler}, {M{\o}ller},
  {Norup S{\o}rensen}, {N{\o}rregaard}, {P{\'e}roux}, {Patat}, {Pena}, {Pragt},
  {Reinero}, {Rigal}, {Riva}, {Roelfsema}, {Royer}, {Sacco}, {Santin},
  {Schoenmaker}, {Spano}, {Sweers}, {Ter Horst}, {Tintori}, {Tromp}, {van
  Dael}, {van der Vliet}, {Venema}, {Vidali}, {Vinther}, {Vola}, {Winters},
  {Wistisen}, {Wulterkens}, \& {Zacchei}}]{Vernet11}
{Vernet}, J., {Dekker}, H., {D'Odorico}, S., {et~al.} 2011, \aap, 536, A105,
  \dodoi{10.1051/0004-6361/201117752}

\bibitem[{{Walther} {et~al.}(2019){Walther}, {O{\~n}orbe}, {Hennawi}, \&
  {Luki{\'c}}}]{Walther19}
{Walther}, M., {O{\~n}orbe}, J., {Hennawi}, J.~F., \& {Luki{\'c}}, Z. 2019,
  \apj, 872, 13, \dodoi{10.3847/1538-4357/aafad1}

\bibitem[{{Wang} {et~al.}(2020){Wang}, {Davies}, {Yang}, {Hennawi}, {Fan},
  {Barth}, {Jiang}, {Wu}, {Mudd}, {Ba{\~n}ados}, {Bian}, {Decarli}, {Eilers},
  {Farina}, {Venemans}, {Walter}, \& {Yue}}]{Wang20}
{Wang}, F., {Davies}, F.~B., {Yang}, J., {et~al.} 2020, \apj, 896, 23,
  \dodoi{10.3847/1538-4357/ab8c45}

\bibitem[{{Weinberg} {et~al.}(1997){Weinberg}, {Hernsquit}, {Katz}, {Croft}, \&
  {Miralda-Escud{\'e}}}]{Weinberg97}
{Weinberg}, D.~H., {Hernsquit}, L., {Katz}, N., {Croft}, R., \&
  {Miralda-Escud{\'e}}, J. 1997, in Structure and Evolution of the
  Intergalactic Medium from QSO Absorption Line System, ed. P.~{Petitjean} \&
  S.~{Charlot}, 133

\bibitem[{{Weinberger} {et~al.}(2019){Weinberger}, {Haehnelt}, \&
  {Kulkarni}}]{Weinberger19}
{Weinberger}, L.~H., {Haehnelt}, M.~G., \& {Kulkarni}, G. 2019, \mnras, 485,
  1350, \dodoi{10.1093/mnras/stz481}

\bibitem[{{Wold} {et~al.}(2022){Wold}, {Malhotra}, {Rhoads}, {Wang}, {Hu},
  {Perez}, {Zheng}, {Khostovan}, {Walker}, {Barrientos},
  {Gonz{\'a}lez-L{\'o}pez}, {Harish}, {Infante}, {Jiang}, {Pharo},
  {Moya-Sierralta}, {Bauer}, {Galaz}, {Valdes}, \& {Yang}}]{Wold22}
{Wold}, I. G.~B., {Malhotra}, S., {Rhoads}, J., {et~al.} 2022, \apj, 927, 36,
  \dodoi{10.3847/1538-4357/ac4997}

\bibitem[{{Worseck} {et~al.}(2014){Worseck}, {Prochaska}, {O'Meara}, {Becker},
  {Ellison}, {Lopez}, {Meiksin}, {M{\'e}nard}, {Murphy}, \&
  {Fumagalli}}]{Worseck14}
{Worseck}, G., {Prochaska}, J.~X., {O'Meara}, J.~M., {et~al.} 2014, \mnras,
  445, 1745, \dodoi{10.1093/mnras/stu1827}

\bibitem[{{Wu} {et~al.}(2021){Wu}, {McQuinn}, \& {Eisenstein}}]{Wu21}
{Wu}, X., {McQuinn}, M., \& {Eisenstein}, D. 2021, \jcap, 2021, 042,
  \dodoi{10.1088/1475-7516/2021/02/042}

\bibitem[{{Wu} {et~al.}(2019){Wu}, {McQuinn}, {Kannan}, {D'Aloisio}, {Bird},
  {Marinacci}, {Dav{\'e}}, \& {Hernquist}}]{Wu19}
{Wu}, X., {McQuinn}, M., {Kannan}, R., {et~al.} 2019, \mnras, 490, 3177,
  \dodoi{10.1093/mnras/stz2807}

\bibitem[{{Wyithe} \& {Bolton}(2011)}]{WB11}
{Wyithe}, J.~S.~B., \& {Bolton}, J.~S. 2011, \mnras, 412, 1926,
  \dodoi{10.1111/j.1365-2966.2010.18030.x}

\bibitem[{{Yang} {et~al.}(2020{\natexlab{a}}){Yang}, {Wang}, {Fan}, {Hennawi},
  {Davies}, {Yue}, {Banados}, {Wu}, {Venemans}, {Barth}, {Bian}, {Boutsia},
  {Decarli}, {Farina}, {Green}, {Jiang}, {Li}, {Mazzucchelli}, \&
  {Walter}}]{Yang20a}
{Yang}, J., {Wang}, F., {Fan}, X., {et~al.} 2020{\natexlab{a}}, \apjl, 897,
  L14, \dodoi{10.3847/2041-8213/ab9c26}

\bibitem[{{Yang} {et~al.}(2020{\natexlab{b}}){Yang}, {Wang}, {Fan}, {Hennawi},
  {Davies}, {Yue}, {Eilers}, {Farina}, {Wu}, {Bian}, {Pacucci}, \&
  {Lee}}]{Yang20b}
---. 2020{\natexlab{b}}, \apj, 904, 26, \dodoi{10.3847/1538-4357/abbc1b}

\bibitem[{{Zel'dovich}(1970)}]{Zel'dovich70}
{Zel'dovich}, Y.~B. 1970, \aap, 5, 84

\bibitem[{{Zhu} {et~al.}(2021){Zhu}, {Becker}, {Bosman}, {Keating},
  {Christenson}, {Ba{\~n}ados}, {Bian}, {Davies}, {D'Odorico}, {Eilers}, {Fan},
  {Haehnelt}, {Kulkarni}, {Pallottini}, {Qin}, {Wang}, \& {Yang}}]{Zhu21}
{Zhu}, Y., {Becker}, G.~D., {Bosman}, S. E.~I., {et~al.} 2021, \apj, 923, 223,
  \dodoi{10.3847/1538-4357/ac26c2}

\bibitem[{{Zhu} {et~al.}(2022){Zhu}, {Becker}, {Bosman}, {Keating},
  {D'Odorico}, {Davies}, {Christenson}, {Ba{\~n}ados}, {Bian}, {Bischetti},
  {Chen}, {Davies}, {Eilers}, {Fan}, {Gaikwad}, {Greig}, {Haehnelt},
  {Kulkarni}, {Lai}, {Pallottini}, {Qin}, {Ryan-Weber}, {Walter}, {Wang}, \&
  {Yang}}]{Zhu22}
---. 2022, \apj, 932, 76, \dodoi{10.3847/1538-4357/ac6e60}

\bibitem[{{Zhu} {et~al.}(2023){Zhu}, {Becker}, {Christenson}, {D'Aloisio},
  {Bosman}, {Bakx}, {D'Odorico}, {Bischetti}, {Cain}, {Davies}, {Davies},
  {Eilers}, {Fan}, {Gaikwad}, {Haehnelt}, {Keating}, {Kulkarni}, {Lai}, {Ma},
  {Mesinger}, {Qin}, {Satyavolu}, {Takeuchi}, {Umehata}, \& {Yang}}]{Zhu23}
{Zhu}, Y., {Becker}, G.~D., {Christenson}, H.~M., {et~al.} 2023, arXiv
  e-prints, arXiv:2308.04614, \dodoi{10.48550/arXiv.2308.04614}

\end{thebibliography}
 \newcommand{\noop}[1]{}

\appendix

\section{Correcting for coarsely-spaced snapshot redshifts}\label{sec:appendix_snaps}

While the data we consider arise from a range of redshifts from $z=5.0$ to $z=6.1$ in steps of $dz=0.1$, we only have outputs from our hydrodynamical simulation in steps of $dz=0.5$. There is thus the potential for biased or incorrect inference of model parameters at redshifts between snapshots.

\begin{figure}
\begin{center}
\resizebox{8cm}{!}{\includegraphics[trim={1.0em 1em 1.0em 1em},clip]{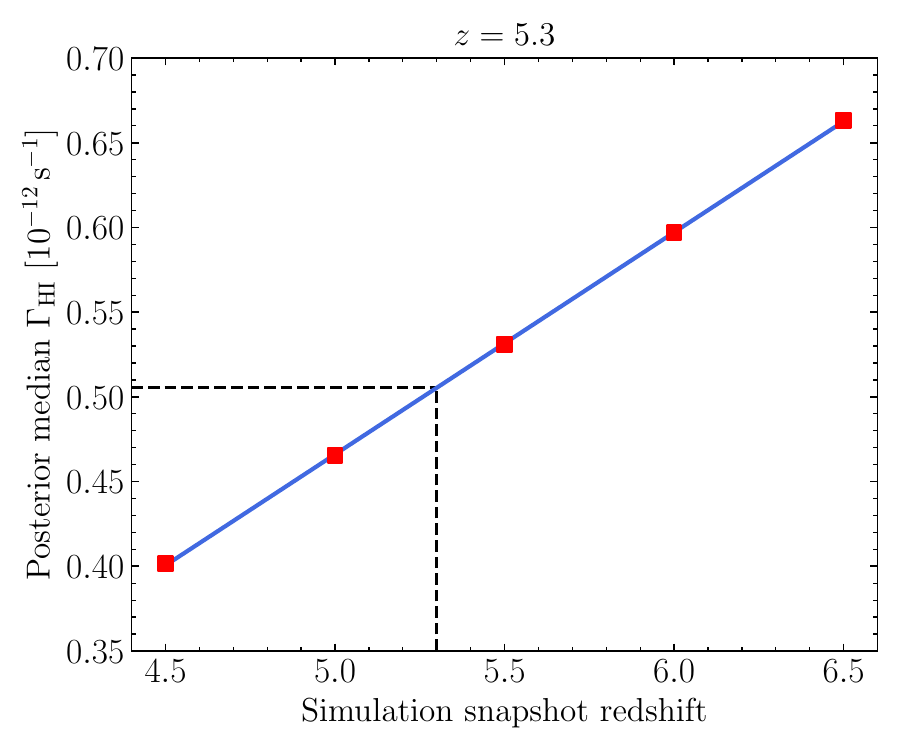}}
\caption{Variation of the posterior median $\Gamma_{\rm HI}$ at $z=5.3$ as a function of hydrodynamical simulation snapshot redshift (red points). The blue curve shows a linear fit to $\Gamma_{\rm HI}$ as a function of snapshot redshift. In this example, the posterior median $\Gamma_{\rm HI}$ at $z=5.3$ is adjusted from $0.532\times10^{-12}$\,s$^{-1}$ to $0.505\times10^{-12}$\,s$^{-1}$.}
\label{fig:snapshots}
\end{center}
\end{figure}

We correct our posterior constraints by performing the full inference procedure described in the text but with five fixed simulation snapshots with redshifts $z_{\rm snap}=$\,4.5, 5.0, 5.5, 6.0, and 6.5. We found that different snapshots resulted in substantially shifted constraints on $\Gamma_{\rm HI}$, with only a minor shift in $\lambda_{\rm mfp}$. For $\Gamma_{\rm HI}$ we found that the shift of the posterior percentiles with snapshot redshift was extremely well described by a linear fit in $\Gamma_{\rm HI}(z_{\rm snap})$, as shown for the posterior median $\Gamma_{\rm HI}$ at $z=5.3$ in Figure~\ref{fig:snapshots}. For our fiducial constraints presented in Figure~\ref{fig:abcghi2} and in Table~\ref{tab:results}, we have shifted the reported posterior percentiles by evaluating the corresponding polynomial fit at the true redshift of the Ly$\alpha$ forest data. These corrections amount to a maximum of $\sim6\%$, and as seen by the very tight relation in Figure~\ref{fig:snapshots}, they contribute a negligible additional source of uncertainty.

\end{document}